\documentclass[aps,prc,superscriptaddress,twocolumn,floatfix]{revtex4}
\usepackage{epsfig}
\usepackage[colorlinks=true,linktocpage=true,linkcolor=blue,citecolor=blue]{hyperref}
\usepackage[usenames,dvipsnames]{color}
\usepackage{amsmath, amssymb}
\usepackage{multirow}
\usepackage{longtable}
\usepackage{color}
\usepackage[normalem]{ulem}  

\renewcommand\sout{\bgroup \color{blue} \ULdepth=-.5ex \ULset}

\newcommand{\I}{{\rm I}}
\newcommand{\T}{{\rm T}}

\begin{document}

\title{Hierarchy of kinetic freeze-out parameters in low energy heavy-ion collisions}
\author{Sudhir Pandurang Rode}
\affiliation{Discipline of Physics, School of Basic Sciences, Indian Institute of Technology Indore, Indore 453552 India}
\author{Partha Pratim Bhaduri}
\affiliation{Variable Energy Cyclotron Centre, HBNI, 1/AF Bidhan Nagar, Kolkata 700 064, India}
\author{Amaresh Jaiswal}
\affiliation{School of Physical Sciences, National Institute of Science Education and Research, HBNI, Jatni-752050, Odisha, India}
\author{Ankhi Roy}
\affiliation{Discipline of Physics, School of Basic Sciences, Indian Institute of Technology Indore, Indore 453552 India}
\date{\today}


\begin{abstract}
We study the mass dependent hierarchy of kinetic freeze-out parameters of hadrons in low energy heavy-ion collisions. For this purpose, the transverse momentum and rapidity spectra of the identified hadrons produced in central Pb+Pb collisions, available at SPS energies ranging from $\rm E_{Lab}=20A-158A $ GeV, are analyzed within a generalized non boost-invariant blast wave model. We consider separate simultaneous fits for light hadrons ($\pi^{-}$, $K^{\pm}$) and heavy strange hadrons ($\Lambda$, $\bar{\Lambda}$, $\phi$, $\Xi^{\pm}$, $\Omega^{\pm}$), for which the transverse momentum spectra as well as rapidity spectra are available. We also perform a separate fit to transverse momentum spectra of charmonia ($J/\Psi$, $\Psi'$) at $158A $ GeV collisions. We find a clear mass dependent hierarchy in the fitted kinetic freeze-out parameters. Further, we study the rapidity spectra using analytical Landau flow solution for non-conformal systems. We find that the fitted value of sound velocity in the medium also shows a similar hierarchy. 
\end{abstract}

\maketitle


\section{Introduction}
\label{}

%
\begin{figure*}[t]
\begin{picture}(160,140)
\put(0,0){\includegraphics[scale=0.28]{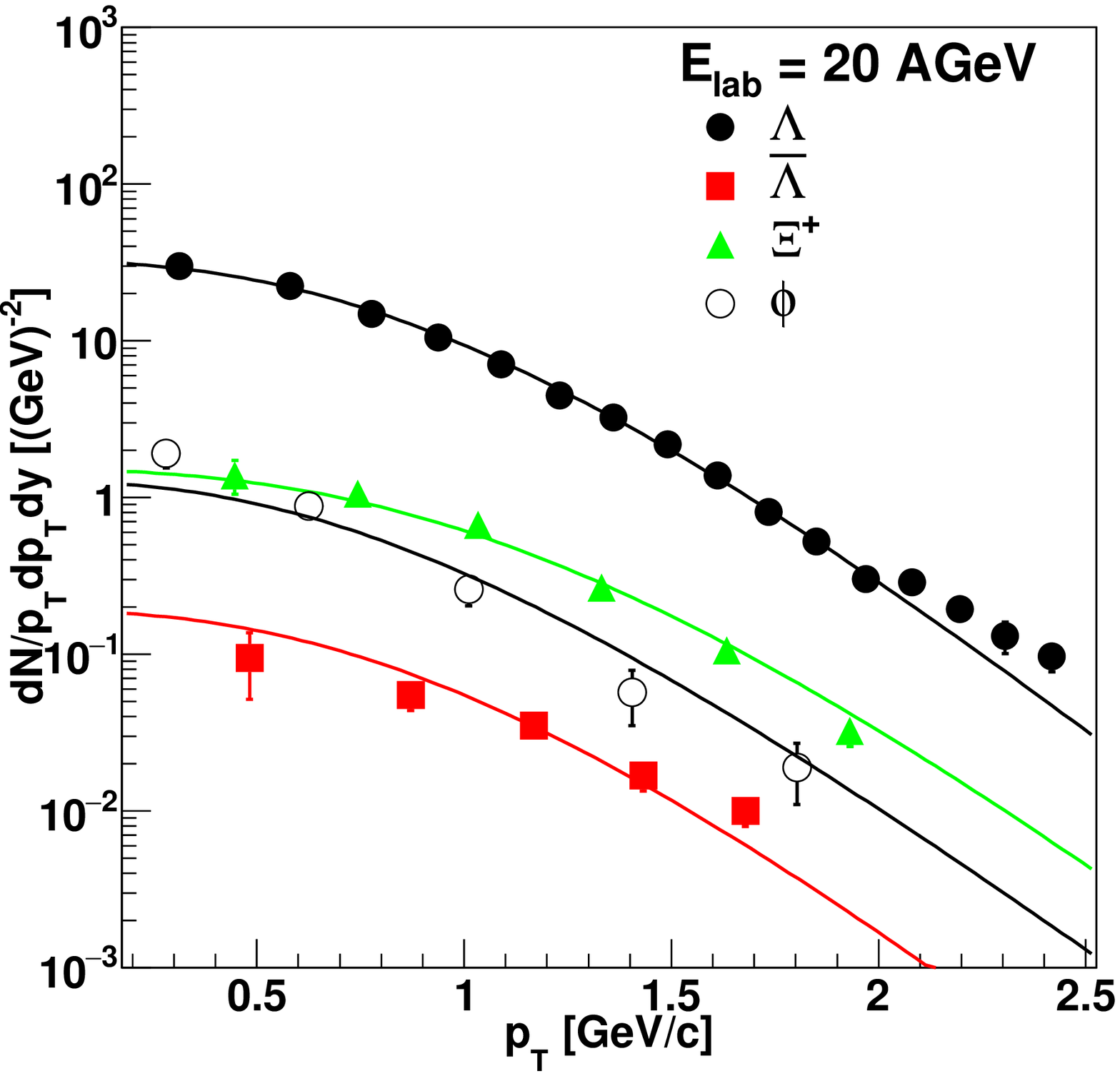}}
\put(50,125){(a)}
\end{picture}
\begin{picture}(160,140)
\put(0,0){\includegraphics[scale=0.28]{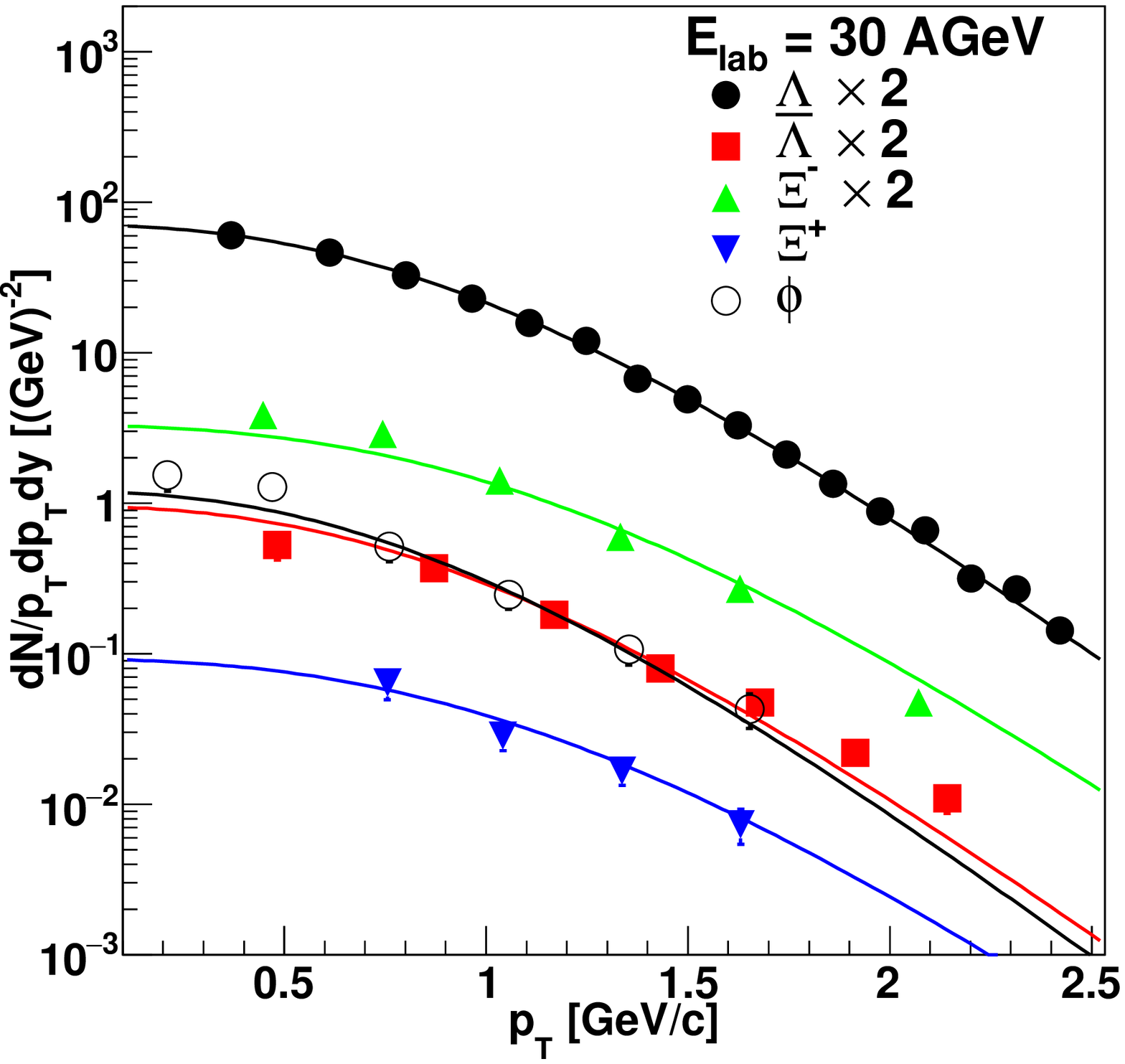}}
\put(50,125){(b)}
\end{picture}
\begin{picture}(160,160)
\put(0,0){\includegraphics[scale=0.28]{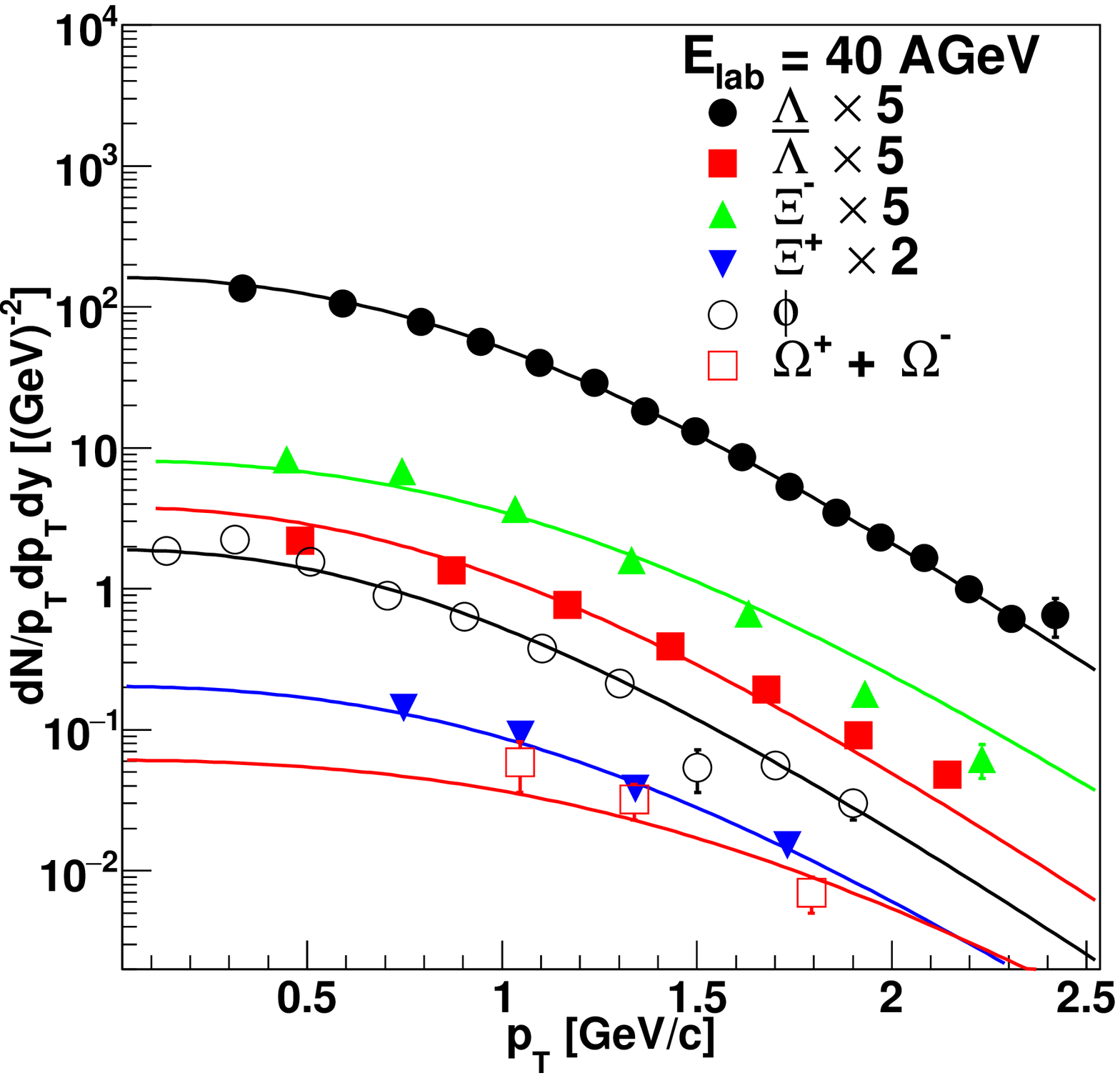}}
\put(50,125){(c)}
\end{picture}
\begin{picture}(160,160)
\put(0,0){\includegraphics[scale=0.28]{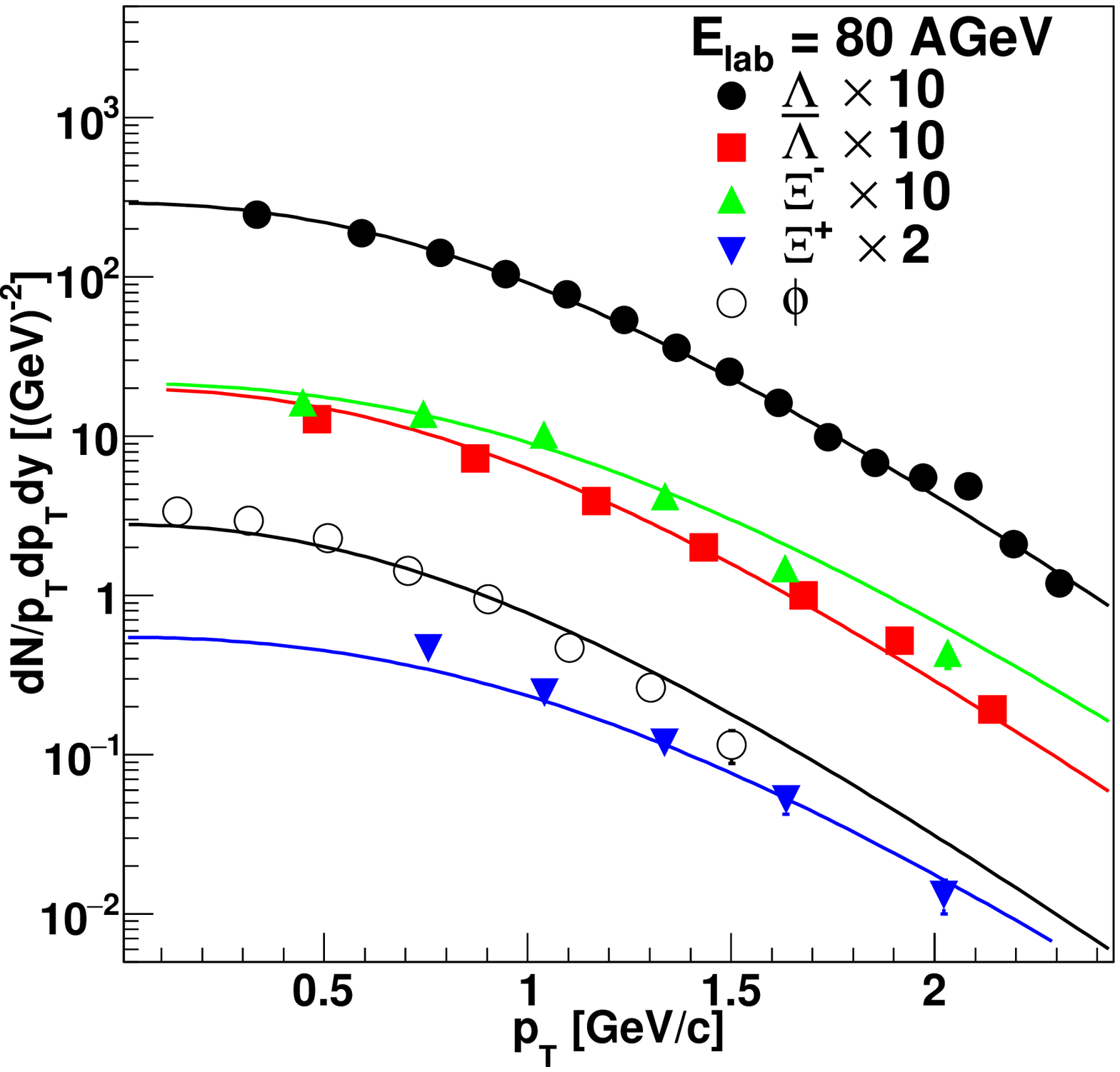}}
\put(50,125){(d)}
\end{picture}
\begin{picture}(160,160)
\put(0,0){\includegraphics[scale=0.28]{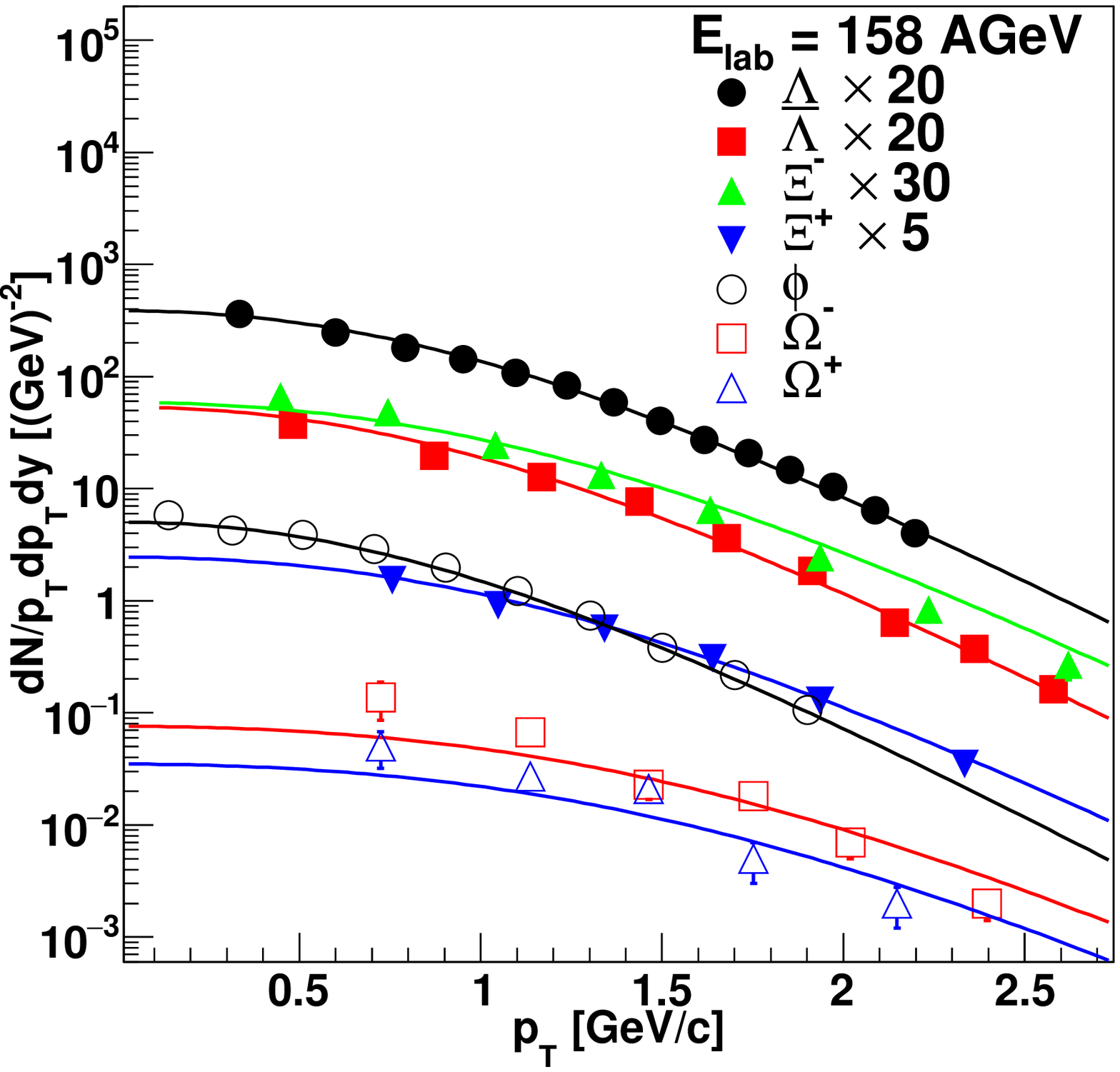}}
\put(50,125){(e)}
\end{picture}
\caption{Simultaneously fitted $p_{T}$ spectra of $\Lambda$, $\bar{\Lambda}$, $\phi$, $\Xi^{\pm}$ and $\Omega^{\pm}$ at (a) 20A GeV, (b) 30A GeV, (c) 40A GeV, (d) 80A GeV and  (e) 158A GeV beam energies.  Error bars indicate available statistical error. 
}
\label{fig4}
\end{figure*}
%

\begin{figure*}
\begin{picture}(160,140)
\put(0,0){\includegraphics[scale=0.28]{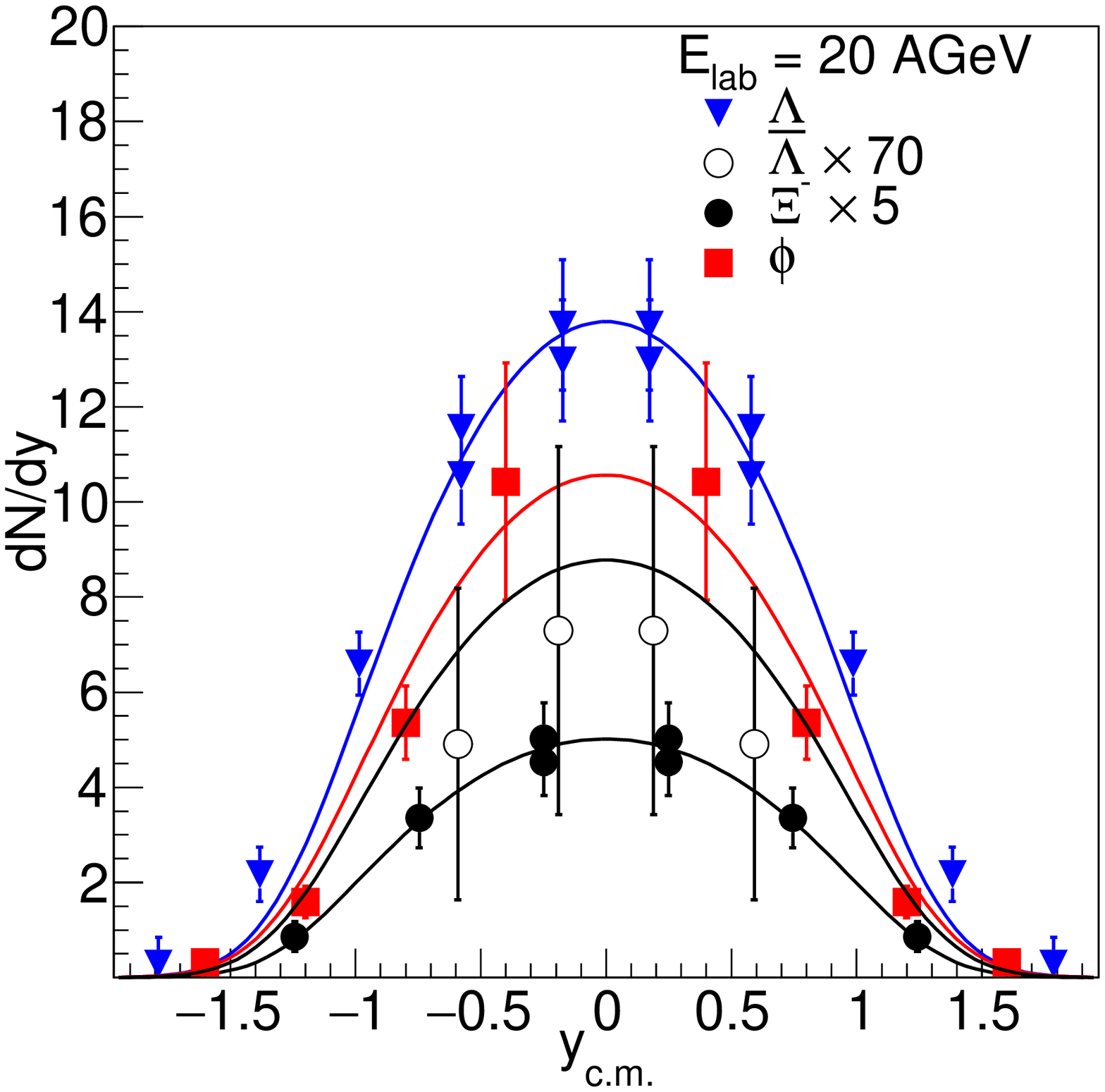}}
\put(50,110){(a)}
\end{picture}
\begin{picture}(160,140)
\put(0,0){\includegraphics[scale=0.28]{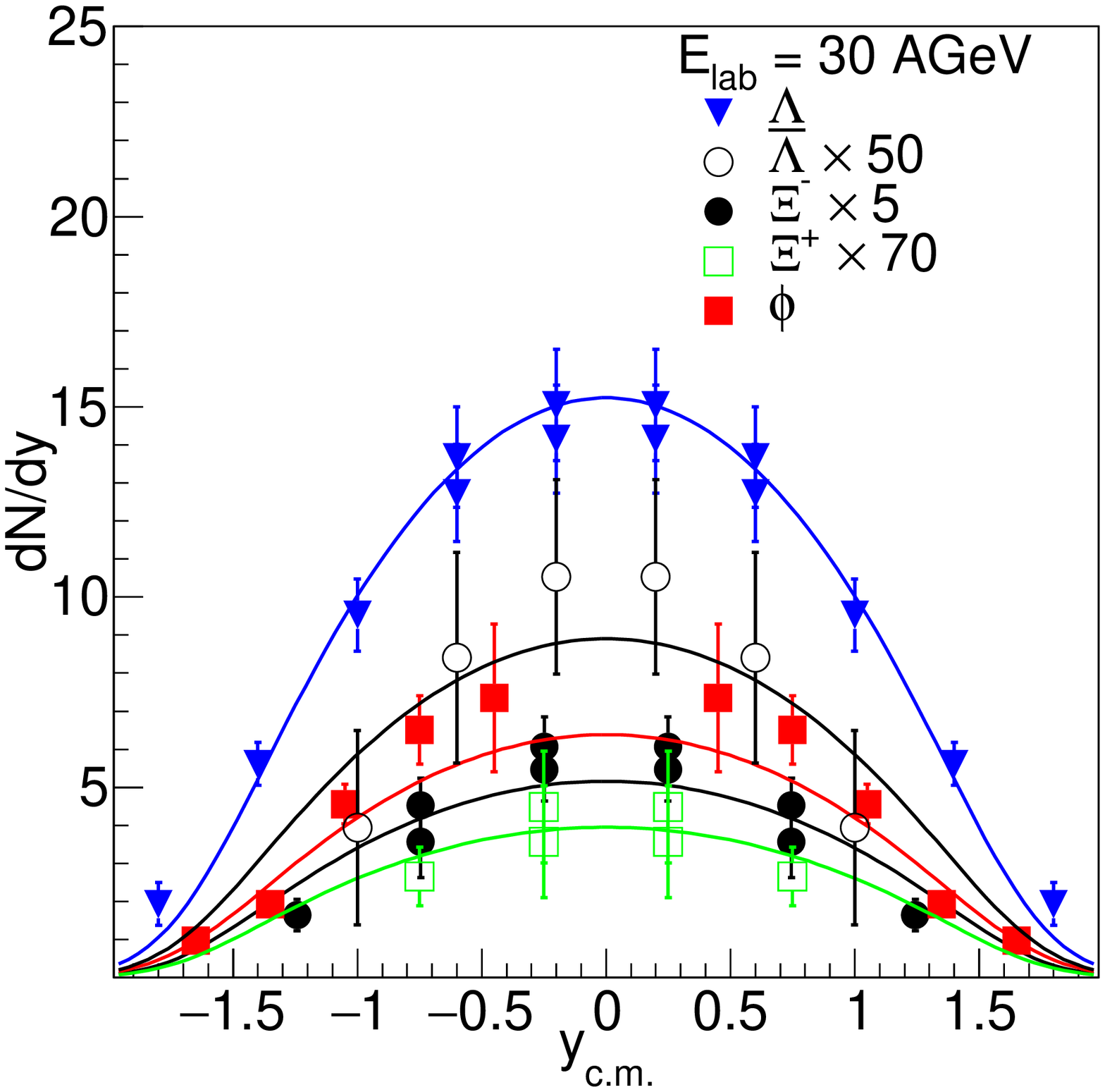}}
\put(50,110){(b)}
\end{picture}
\begin{picture}(160,160)
\put(0,0){\includegraphics[scale=0.28]{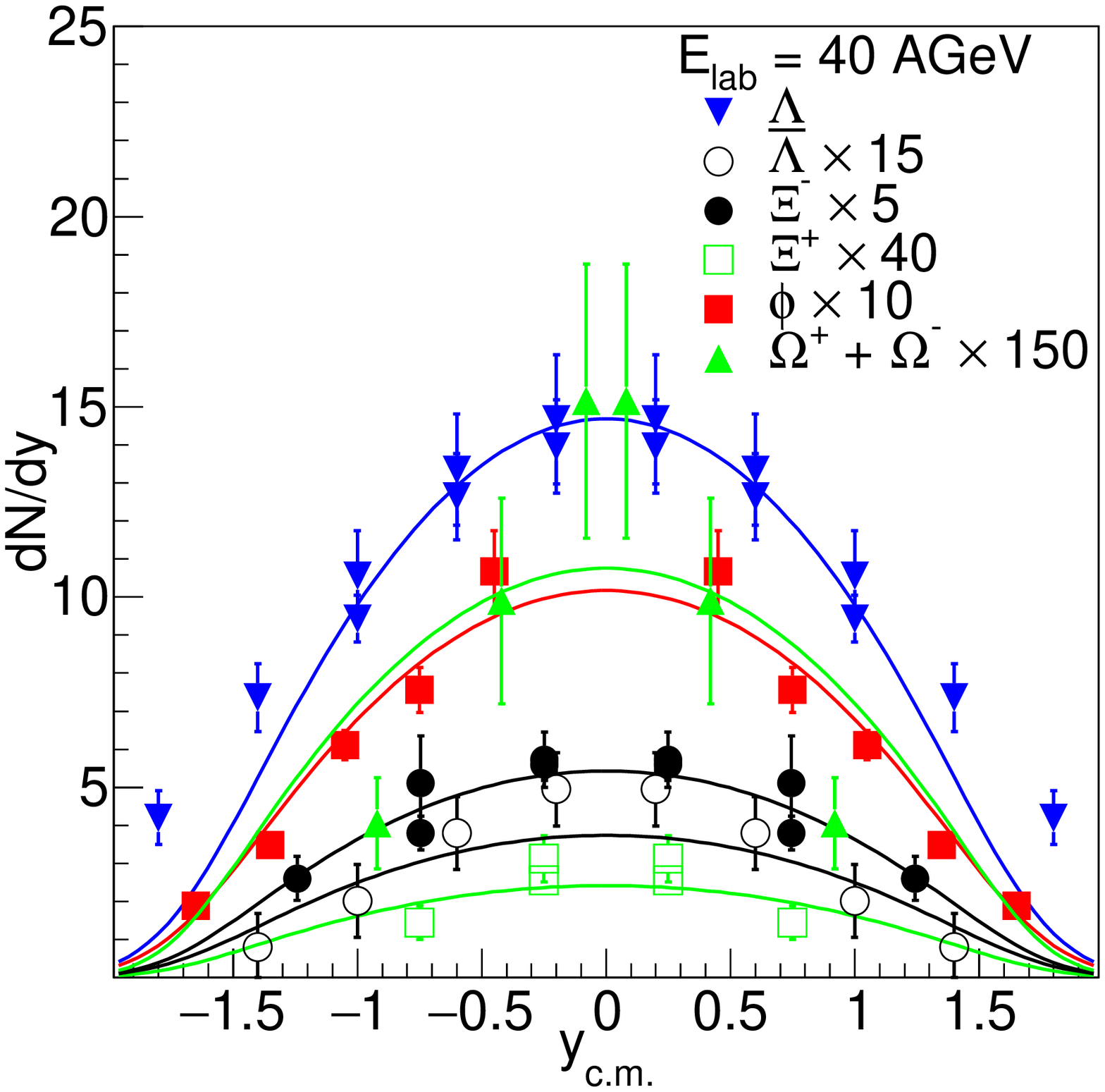}}
\put(50,110){(c)}
\end{picture}
\begin{picture}(160,160)
\put(0,0){\includegraphics[scale=0.28]{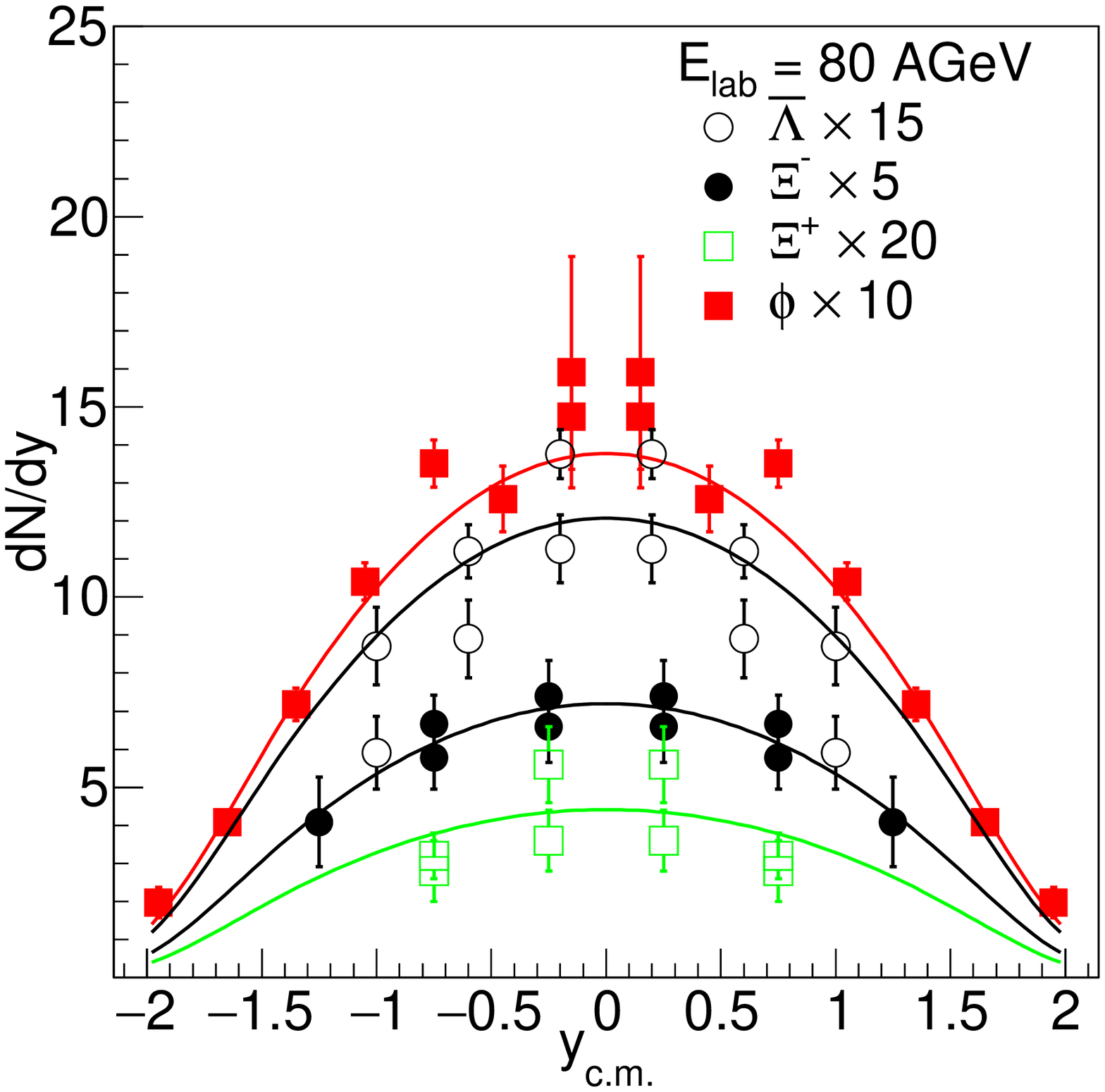}}
\put(50,110){(d)}
\end{picture}
\begin{picture}(160,160)
\put(0,0){\includegraphics[scale=0.28]{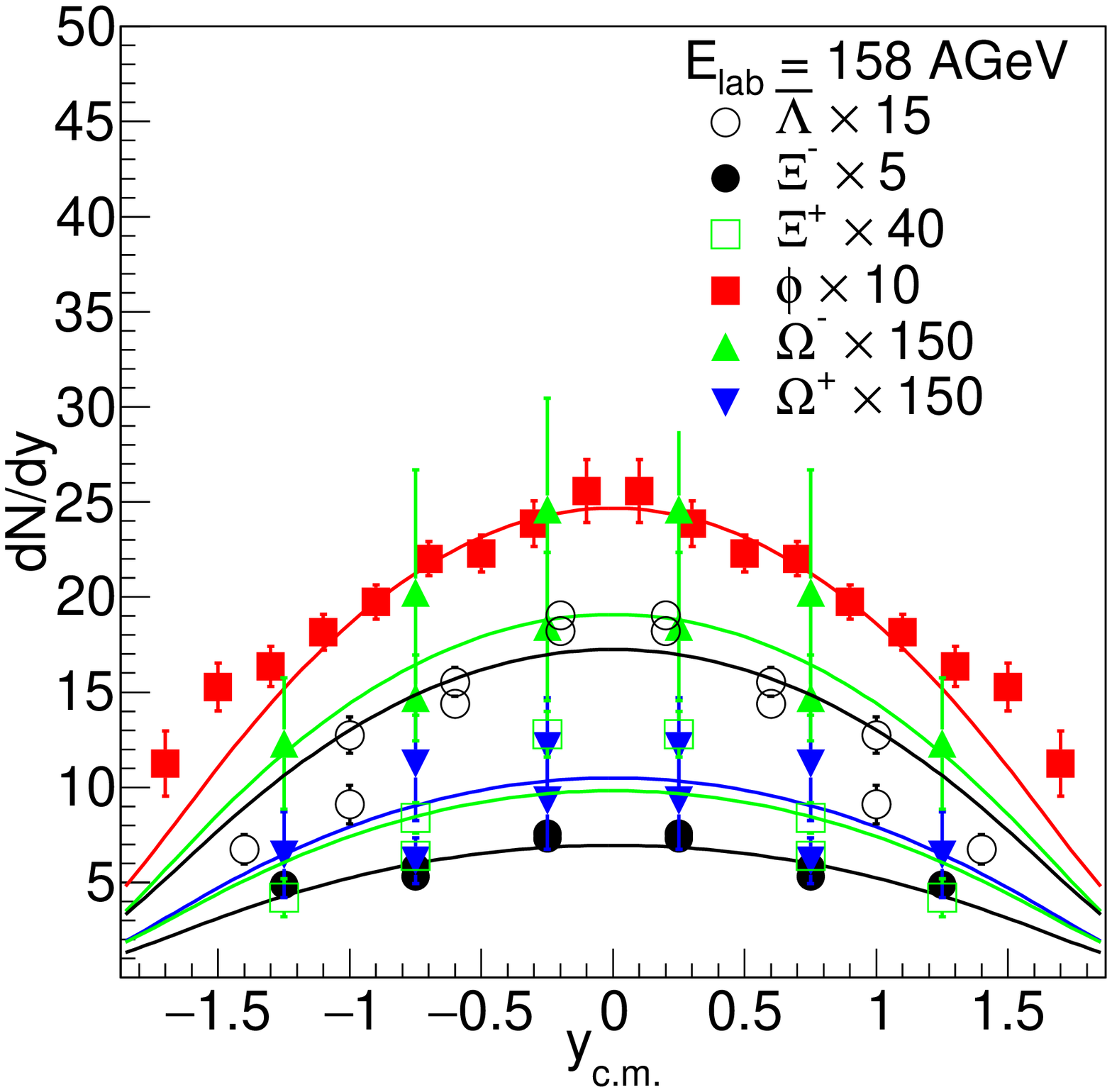}}
\put(50,110){(e)}
\end{picture}
\caption{Simultaneously fitted rapidity distribution of $\Lambda$, $\bar{\Lambda}$, $\phi$, $\Xi^{\pm}$ and $\Omega^{\pm}$ in central Pb+Pb collisions from SPS, at (a) 20A GeV, (b) 30A GeV, (c) 40A GeV, (d) 80A GeV and  (e) 158A GeV beam energies. Error bars indicate available statistical error.}
\label{fig7}
\end{figure*}
Relativistic heavy-ion collisions are a suitable tool to produce and study hot and dense strongly interacting matter in the laboratory~\cite{Florkowski:2014yza, UWHeinz, BraunMunzinger:2008tz}. By varying the collision energy, the nuclear matter can be created over a wide range of temperatures and densities, which facilitates the systematic investigation of the large parts of the QCD phase diagram. The ultra-relativistic nuclear collisions at Relativistic Heavy Ion Collider (RHIC)~\cite{rhic1,rhic2} and Large Hadron Collider (LHC)~\cite{lhc1,lhc2,lhc3} predominantly produces a partonic medium at high temperature and vanishingly small baryon chemical potential, thermodynamic properties of which are most suitably studied using lattice QCD (lQCD) simulations \cite{Fodor:2009ax, Bazavov:2009zn, Borsanyi:2010cj, Bazavov:2014pvz, Borsanyi:2013bia}.

In relativistic nuclear collisions at lower energies, nuclear matter is created at high net baryon densities and moderate temperature, where scope of application of lQCD is rather limited. On experimental front, this has led to renewed interest in collision at moderate energies, manifested in the ongoing and upcoming experimental programs at RHIC~\cite{Bzdak:2019pkr}, SPS~\cite{Lewicki:2020mqr,Agnello:2018evr}, FAIR~\cite{CBM-physics,Senger:2019ccc} and NICA~\cite{nica}. An optimum utilization of the future facilities demands a coherent interpretation of the available data sets in the similar energy range collected by the previous generation fixed target experiments at AGS and SPS accelerator facilities. Of particular interest is the determination of freeze-out conditions of the fireball at various collision energies.

During chemical freeze-out the inelastic scatterings cease, leading to the stabilization of the particle chemistry in the fireball. On the other hand, at kinetic or thermal freeze-out hadrons stop to interact with each other and their momentum distribution does not undergo further change. In the so called ``standard model'' of heavy-ion collisions, chemical freeze-out occurs earlier than kinetic freeze-out due to larger mean free path of inelastic collisions~\cite{Heinz:2007in}. Usually, the yields and transverse momentum ($p_T$) spectra of the produced hadrons are analyzed to extract the parameters of chemical and kinetic freeze-out. In Ref.~\cite{Chatterjee:2013yga} the authors advocated for a multiple chemical freeze-out scenario, with strange hadrons fixing their chemical composition earlier than the non-strange light hadrons, due to smaller inelastic cross-sections. An interesting question to ask is whether a similar hierarchical structure is also present in the case of kinetic decoupling. One may expect a mass dependent hierarchy of kinetic freeze-out as the medium induced momentum change of heavy hadrons would be smaller compared to lighter hadrons. Therefore, as the temperature of the fireball decreases, one would expect an earlier kinetic decoupling of heavy hadrons. In the present article, we have made an attempt to look for the possible hierarchy in thermal freeze-out, in low energy nuclear collisions.

Kinetic freeze-out conditions are commonly studied in hydrodynamics inspired blast-wave model framework~\cite{Florkowski:2004tn}. In our previous work~\cite{Rode:2018hlj}, we investigated the kinetic freeze out conditions of the light (bulk) hadrons ($\pi^{-}$, $K^{\pm}$, $p$) by analyzing their transverse and longitudinal spectra in the beam energy range of $E_{Lab} = 2A - 158A$ GeV by employing a non boost-invariant blast-wave model formulated in Ref.~\cite{Dobler:1999ju}. In the original blast wave model, the hydrodynamical results for particle spectra are approximated by emission from a cylindrically symmetric and longitudinally boost-invariant fireball \cite{Schnedermann}. In the non boost-invariant extension, the symmetry is explicitly broken by introducing a dependence of the transverse size of the fireball on the space-time rapidity. This is particularly useful for low energy collisions where the longitudinal boost invariance is absent in the measured rapidity ($y$) distributions of the hadrons.

In this article, we employ non boost-invariant blast-wave model to study the mass dependent hierarchy in kinetic freeze-out parameters of hadrons produced in central Pb+Pb collisions at SPS energies. To this end, we analyze the $p_T$-spectra and rapidity spectra of the identified hadrons at collision energies ranging from $\rm E_{Lab}=20A-158 $A GeV. We consider separate simultaneous fits for light hadrons ($\pi^{-}$, $K^{\pm}$) and heavy strange hadrons ($\Lambda$, $\bar{\Lambda}$, $\phi$, $\Xi^{\pm}$, $\Omega^{\pm}$), for which the transverse momentum spectra, as well as rapidity spectra, are available. We do not consider protons in the fits as the rapidity spectra of protons are not available at SPS energies. For heavy strange hadrons, our analysis results indicate a relatively low kinetic freeze-out temperature in the range of $90 - 110$ MeV, with a rather strong mean transverse velocity of collective expansion of about $0.4c - 0.5c$. We also perform a separate fit to transverse momentum spectra of charmed hadrons ($J/\Psi$, $\Psi'$) at $158 $A GeV collisions. We find a clear mass dependent hierarchy in the fitted kinetic freeze-out parameters. Further, we study the rapidity spectra using analytical Landau flow solution for non-conformal systems. We find that the fitted value of sound velocity in the medium also shows a similar hierarchy.

In the present work, we perform for the first time, a systematic analysis of the heavy strange hadrons produced in the low energy nuclear collisions using non boost-invariant blast-wave model. Note that the application of blast-wave dynamics to study the transverse spectra of heavy hadrons have been attempted earlier. In Ref.~\cite{Gorenstein}, the authors have analyzed the $p_T$ spectra of J/$\psi$, $\psi'$ mesons and $\Omega$ baryon within the longitudinal boost-invariant blast-wave model, with the hypothesis that for these heavy hadrons, the rescattering effects in the hadronic phase is negligible and they leave the fireball at hadronization. However, to the best of our knowledge, a thorough analysis of $p_T$ and $y$ distributions of all varieties of multi-strange hadrons produced in the low energy domain has never been attempted before using a non boost-invariant blast wave model.

%
\begin{table*}[t]\centering
\label{tabI}
\vglue4mm
\begin{tabular}{cccccccc} \hline
Facility & Experiment & E$_{\rm Lab}$~(A GeV) & $y_{beam}$ & System & Centrality & Phase space & Hadron Species\\ \hline
\hline
SPS \rule{0pt}{0.5cm} & NA49 & 20 & 3.75 & Pb+Pb &  $0 - 7.2 \%$&$0.0 < y_{\rm c.m.} < 1.8$ ($\phi$) & $\Lambda$ ($\bar{\Lambda}$)~\cite{Alt:2008qm}, $\phi$~\cite{Alt:2008iv}, $\Xi^{\pm}$~\cite{Alt:2008qm}\\
\rule{0pt}{0.5cm} &  &  &  &  &  $0 - 7 \%$&$-0.4 < y_{\rm c.m.} < 0.4$ ($\Lambda$, $\bar{\Lambda}$) & \\
 \rule{0pt}{0.5cm} &  &  &  &  &  $0 - 7 \%$&$-0.5 < y_{\rm c.m.} < 0.5$ ($\Xi^{\pm}$) & \\
\hline
SPS \rule{0pt}{0.5cm} & NA49 & 30 & 4.16 & Pb+Pb & $0 - 7.2 \%$&$0.0 < y_{\rm c.m.} < 1.8$ ($\phi$) & $\Lambda$ ($\bar{\Lambda}$)~\cite{Alt:2008qm}, $\phi$~\cite{Alt:2008iv}, $\Xi^{\pm}$~\cite{Alt:2008qm}\\ 
\rule{0pt}{0.5cm} &  &  &  &  &  $0 - 7 \%$&$-0.4 < y_{\rm c.m.} < 0.4$ ($\Lambda$, $\bar{\Lambda}$) & \\
 \rule{0pt}{0.5cm} &  &  &  &  & $0 - 7 \%$ &$-0.5 < y_{\rm c.m.} < 0.5$ ($\Xi^{\pm}$) & \\
\hline
SPS \rule{0pt}{0.5cm} & NA49 & 40 & 4.45 & Pb+Pb & $0 - 7 \%$&$-0.4 < y_{\rm c.m.} < 0.4$ ($\Lambda$, $\bar{\Lambda}$)& $\Lambda$ ($\bar{\Lambda}$)~\cite{Alt:2008qm}, $\phi$~\cite{Alt:2008iv}, $\Omega^{\pm}$~\cite{Alt:2004kq}, $\Xi^{\pm}$~\cite{Alt:2008qm}\\
\rule{0pt}{0.5cm} &  &  &  &  &  $0 - 7 \%$ &$-0.5 < y_{\rm c.m.} < 0.5$ ($\Xi^{\pm}$) & \\
 \rule{0pt}{0.5cm} &  &  &  &  &  $0 - 7.2 \%$&$0.0 < y_{\rm c.m.} < 1.5$ ($\phi$) & \\
 \rule{0pt}{0.5cm} &  &  &  &  &  $0 - 7.2 \%$ &$-0.5 < y_{\rm c.m.} < 0.5$ ($\Omega^{\pm}$) & \\\hline
SPS \rule{0pt}{0.5cm} & NA49 & 80 & 5.12 & Pb+Pb & $0 - 7 \%$&$-0.4 < y_{\rm c.m.} < 0.4$ ($\Lambda$, $\bar{\Lambda}$)& $\Lambda$ ($\bar{\Lambda}$)~\cite{Alt:2008qm}, $\phi$~\cite{Alt:2008iv}, $\Xi^{\pm}$~\cite{Alt:2008qm}\\
 \rule{0pt}{0.5cm} &  &  &  &  &  $0 - 7.2 \%$ &$0.0 < y_{\rm c.m.} < 1.7$ ($\phi$) & \\
\rule{0pt}{0.5cm} &  &  &  &  &  $0 - 7 \%$ &$-0.5 < y_{\rm c.m.} < 0.5$ ($\Xi^{\pm}$) & \\\hline
SPS \rule{0pt}{0.5cm} & NA49 & 158 & 5.82 & Pb+Pb & $0 - 10 \%$&$-0.4 < y_{\rm c.m.} < 0.4$ ($\Lambda$, $\bar{\Lambda}$) & $\Lambda$ ($\bar{\Lambda}$)~\cite{Alt:2008qm}, $\phi$~\cite{Alt:2008iv}, $\Omega^{\pm}$~\cite{Alt:2004kq}, $\Xi^{\pm}$~\cite{Alt:2008qm}\\
\rule{0pt}{0.5cm} &  &  &  &  & $0 - 5 \%$ &$0.0 < y_{\rm c.m.} < 1.0$ ($\phi$) & \\
\rule{0pt}{0.5cm} &  &  &  &  &$0 - 23.5 \%$ &$-0.5 < y_{\rm c.m.} < 0.5$ ($\Omega^{\pm}$) & \\
\rule{0pt}{0.5cm} &  &  &  &  & $0 - 10 \%$ &$-0.5 < y_{\rm c.m.} < 0.5$ ($\Xi^{\pm}$) & \\\hline
\end{tabular}
\caption{Details of the data sets from different experiments at different accelerator facilities along with energy ($\rm E_{Lab}$), beam rapidity ($y_{beam}$) in lab frame, System, Centrality, Phase space and Hadron species, used for this blast wave analysis.}
\end{table*}
%

The paper is organized as follows. In section II, the essential features of the non boost-invariant model are described in a nutshell. The analysis results are presented in section III. In section IV we summarize our main results and conclude.


\section{A brief description of the model}

%
\begin{figure*}[t]
\begin{picture}(160,140)
\put(0,0){\includegraphics[scale=0.28]{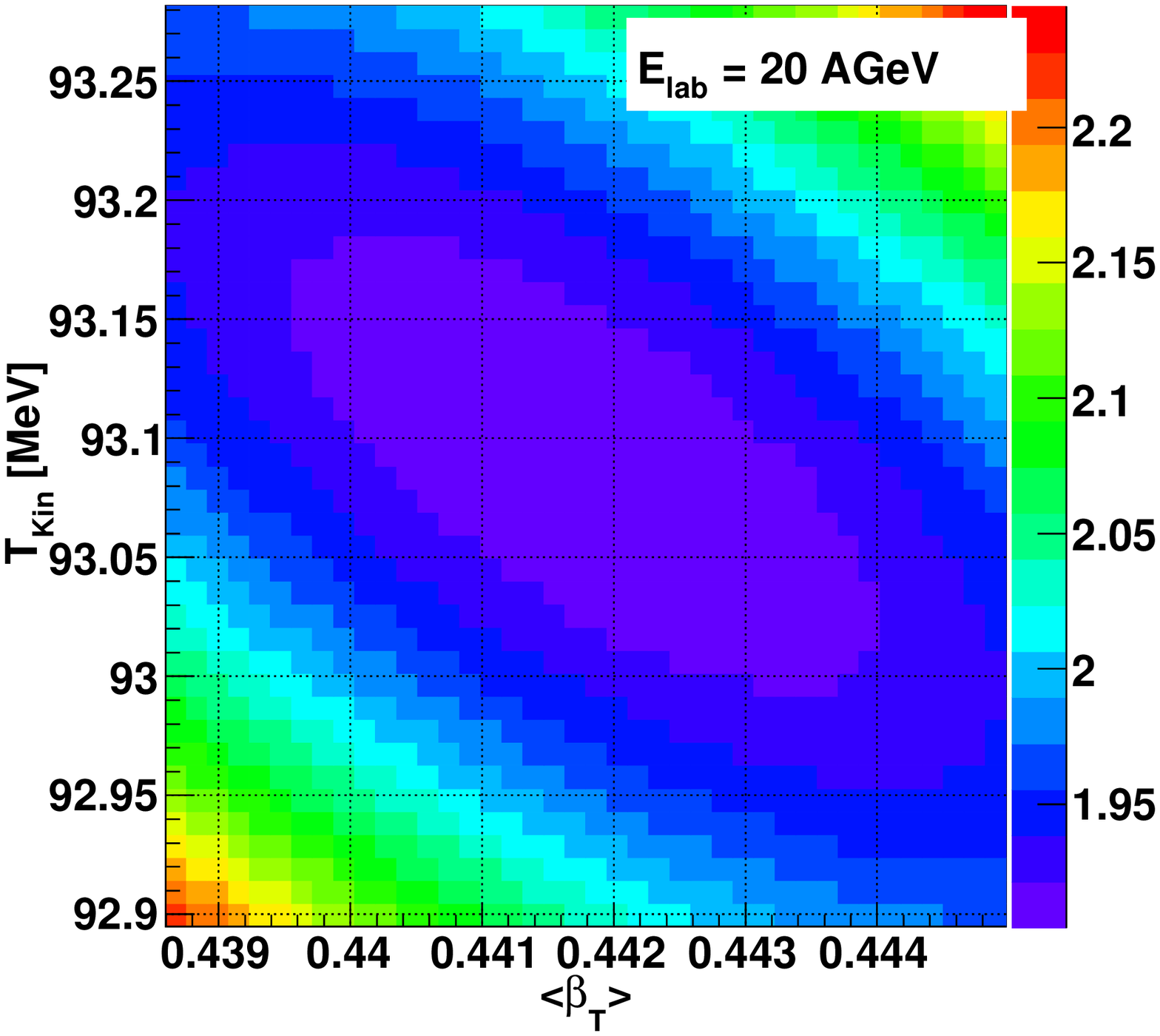}}
\put(50,125){(a)}
\end{picture}
\begin{picture}(160,140)
\put(0,0){\includegraphics[scale=0.28]{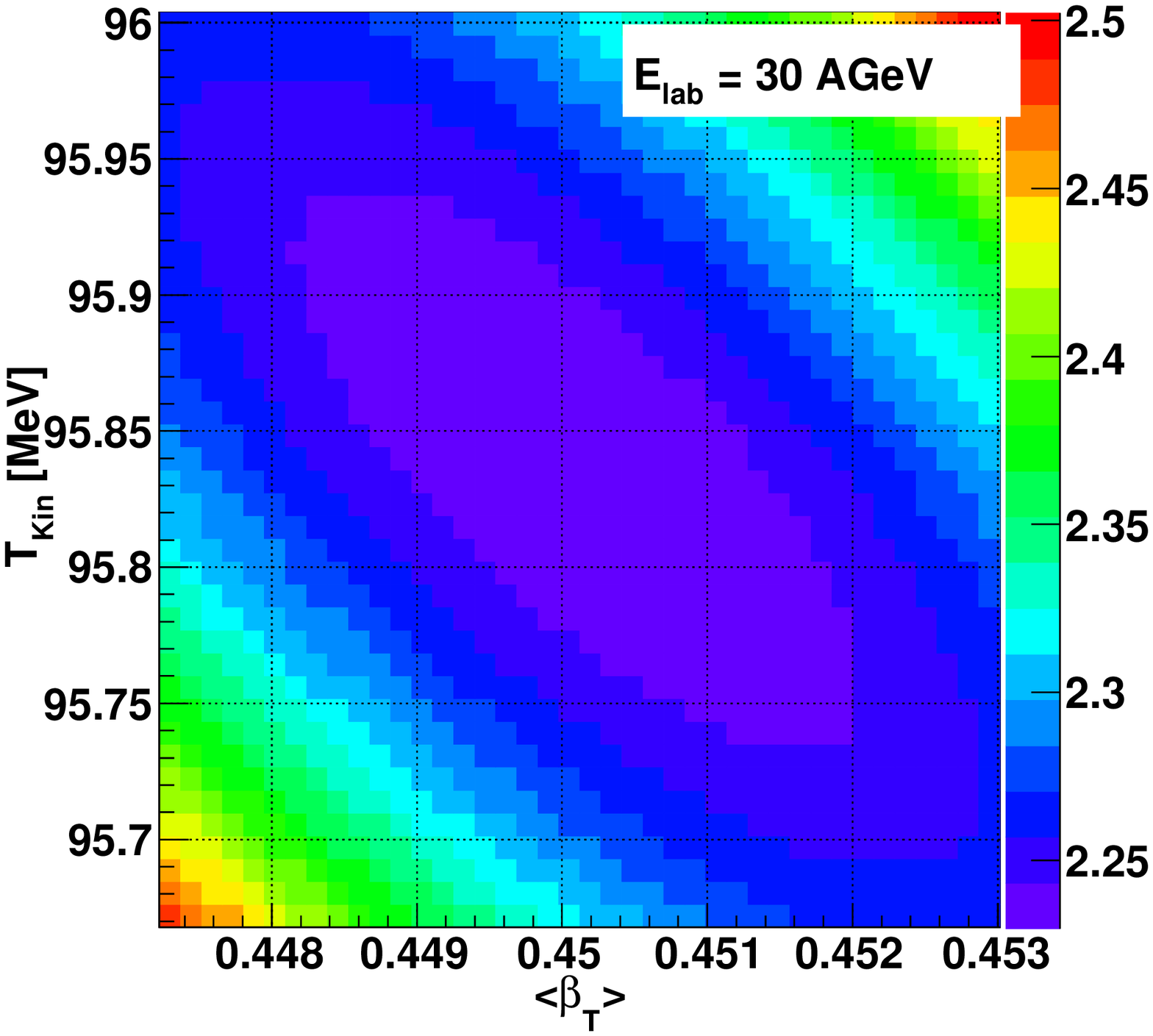}}
\put(50,125){(b)}
\end{picture}
\begin{picture}(160,160)
\put(0,0){\includegraphics[scale=0.28]{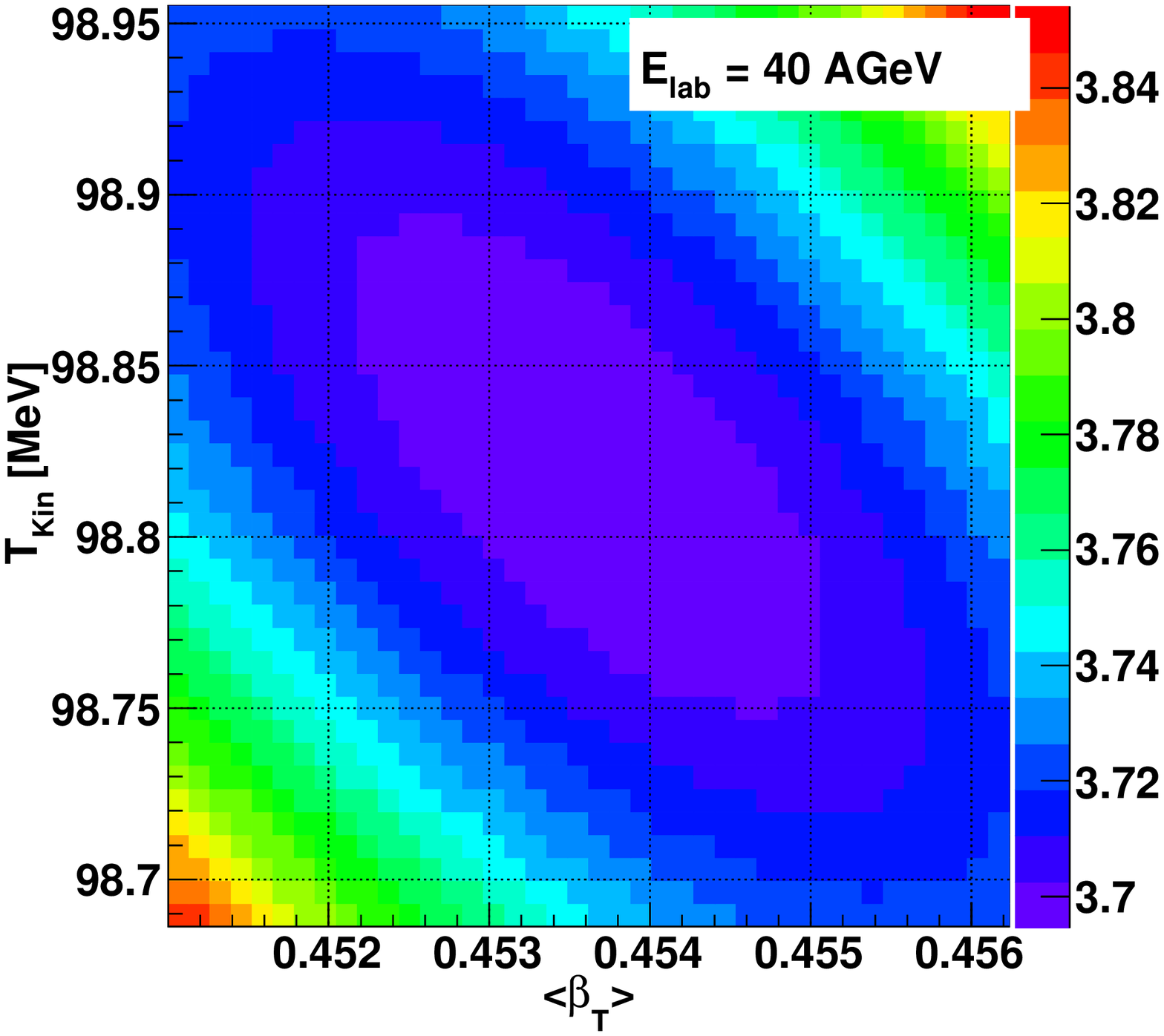}}
\put(50,125){(c)}
\end{picture}
\begin{picture}(160,160)
\put(0,0){\includegraphics[scale=0.28]{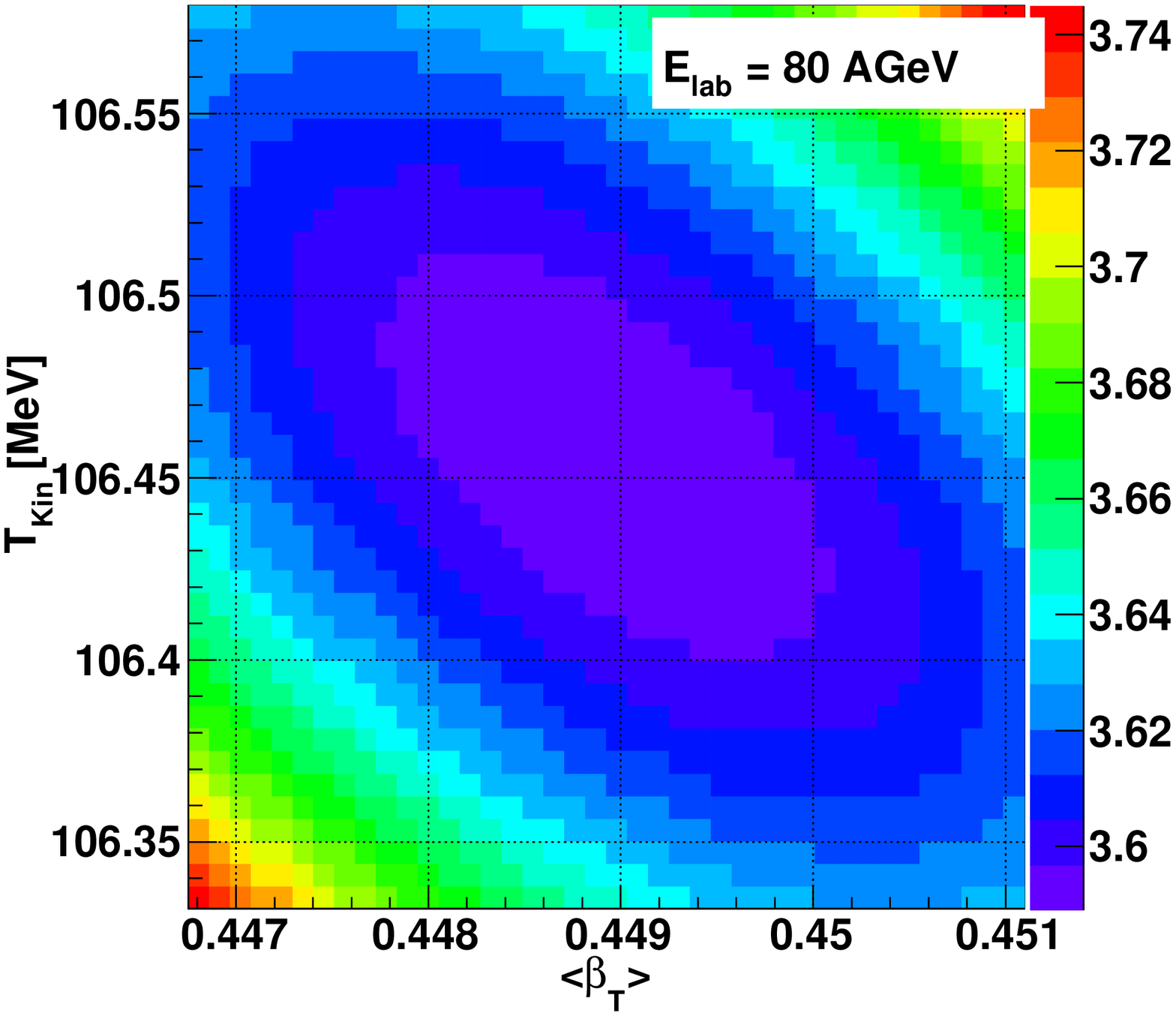}}
\put(50,125){(d)}
\end{picture}
\begin{picture}(160,160)
\put(0,0){\includegraphics[scale=0.28]{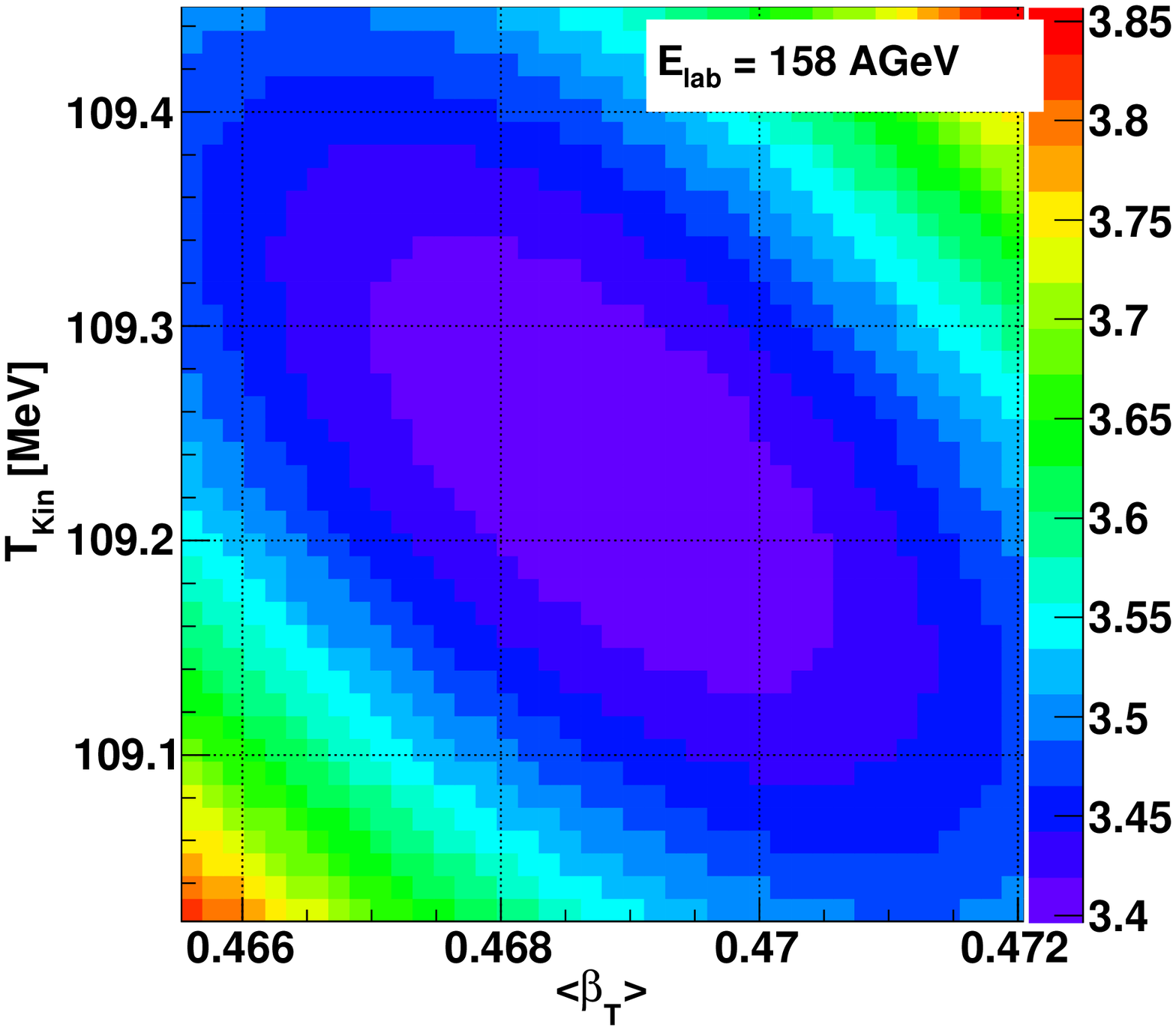}}
\put(50,125){(e)}
\end{picture}
\caption{Two dimensional projections of the $\chi^{2}$ contour plots in the $\langle\beta_T\rangle - T_{\rm kin}$ plane for (a) 20A GeV, (b) 30A GeV, (c) 40A GeV, (d) 80A GeV and  (e) 158A GeV beam energies. Different colors corresponds to different values of $\chi^{2}/N_{\rm dof}$. 
}
\label{contour}
\end{figure*}
%

%
\begin{table*}[ht]\centering
\vglue4mm
\begin{tabular}{ccccccc} \hline
 $\rm E_{Lab}$~(A GeV) & $\eta_{max}$   &$\langle \beta_{T} \rangle$ & $T_{kin}$ (MeV) & $\chi^{2}/N_{\rm dof}$\\ \hline
20 \rule{0pt}{0.5cm} &$1.288 \pm 0.021$&$0.4418 \pm 0.0032$ & $93.09 \pm 0.19$ & 1.90\\\hline
30 \rule{0pt}{0.5cm} &$1.728 \pm 0.026$ &$0.4501 \pm 0.0029$ & $95.84 \pm 0.17$ & 2.23\\\hline
40 \rule{0pt}{0.5cm} & $1.752 \pm 0.018$ &$0.4536 \pm 0.0026$ & $98.82 \pm 0.14$ & 3.70\\
\hline
80 \rule{0pt}{0.5cm} &$1.989 \pm 0.021$ &$0.4489 \pm 0.0022$ & $106.46 \pm 0.12$ & 3.59\\
\hline
158 \rule{0pt}{0.5cm} &$2.031 \pm 0.029$ &$0.4688 \pm 0.0016$ & $109.24 \pm 0.11$ & 3.40\\
\hline
\end{tabular}
\caption{Summary of the fit results of $p_{\rm T}$ spectra of heavy strange hadrons at different energies ranging from 20A to 158A GeV at SPS.}
\label{tabII}
\end{table*}
%

%
\begin{figure*}[ht]
\includegraphics[scale=0.28]{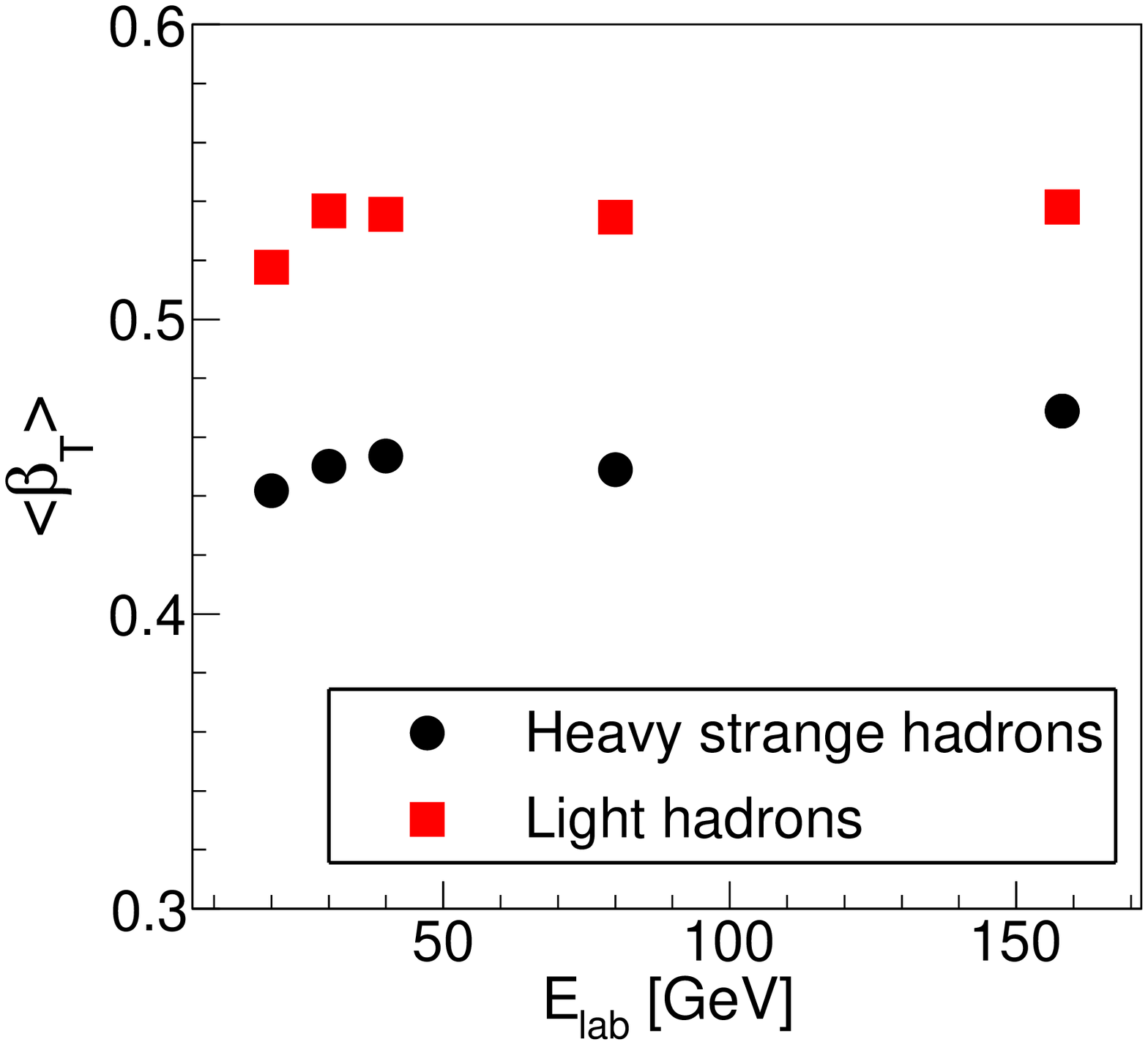}
\includegraphics[scale=0.28]{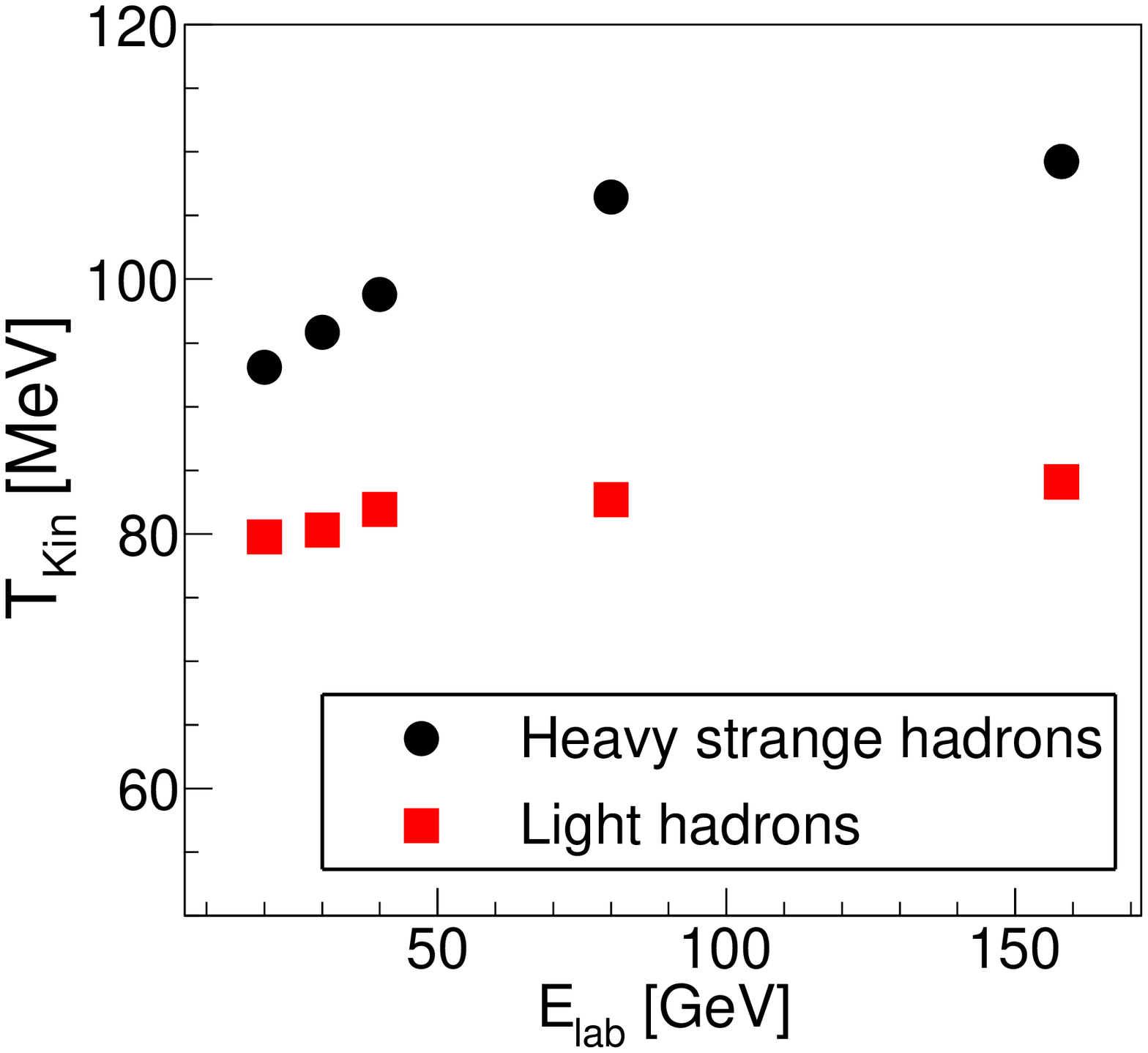}
\includegraphics[scale=0.28]{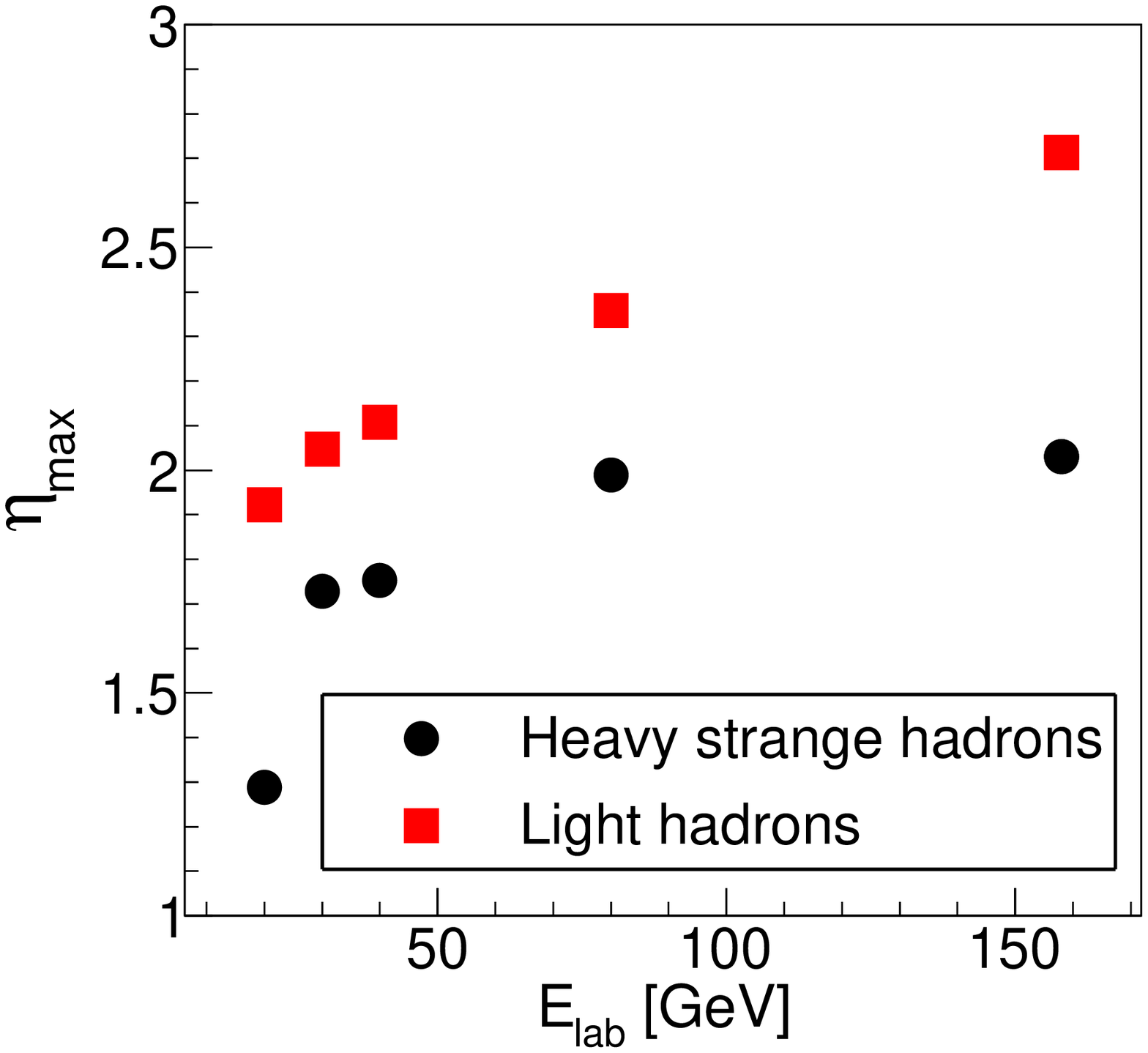}
\caption{Variation of the $\langle \beta_{T} \rangle$ (left), $T_{kin}$ (middle) and  $\eta_{max}$ (right) for heavy strange and light hadrons with incident beam energy ($\rm E_{\rm lab}$). Errors are within the marker size.}
\label{fig9}
\end{figure*}
%

%
\begin{figure}[h!]
\includegraphics[scale=0.4]{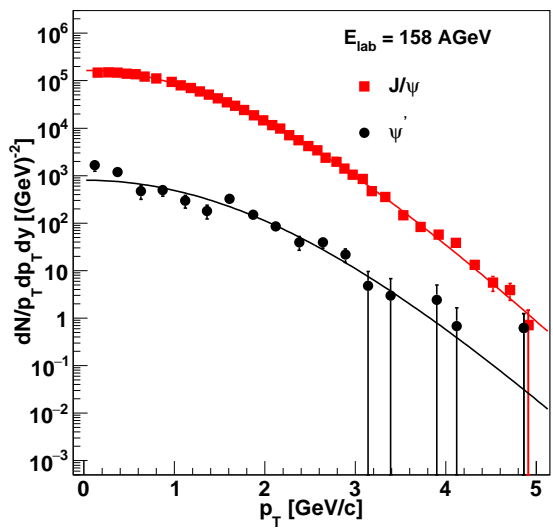}
\caption{Simultaneously fitted $p_{T}$ spectra of J/$\Psi$ and $\Psi^{'}$ at 158A GeV. Uncorrelated statistical and systematic errors are added in quadrature.}
\label{JPsi158AGeVpt}
\end{figure}
%

%
\begin{figure*}[ht]
\includegraphics[scale=0.28]{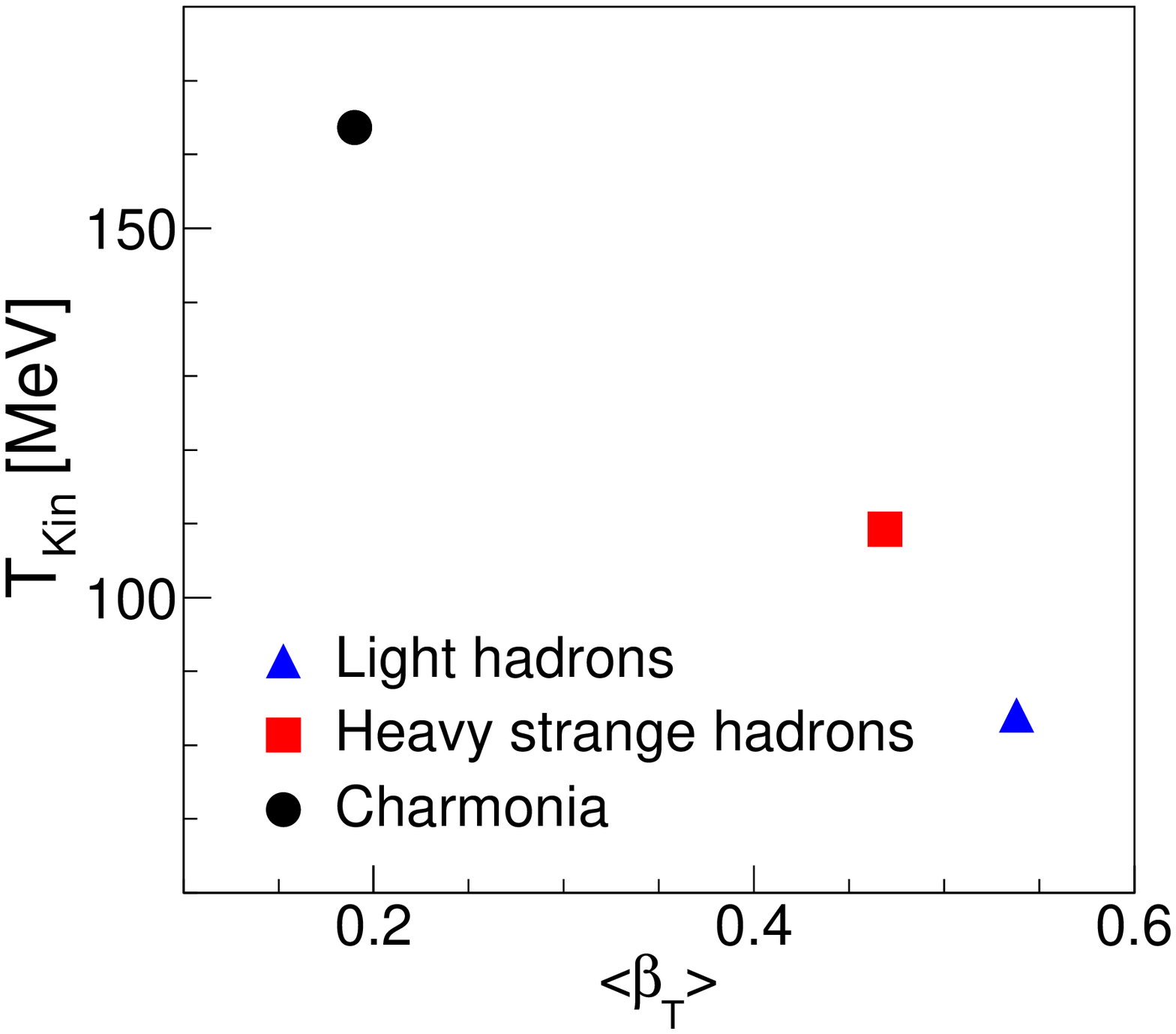}
\includegraphics[scale=0.28]{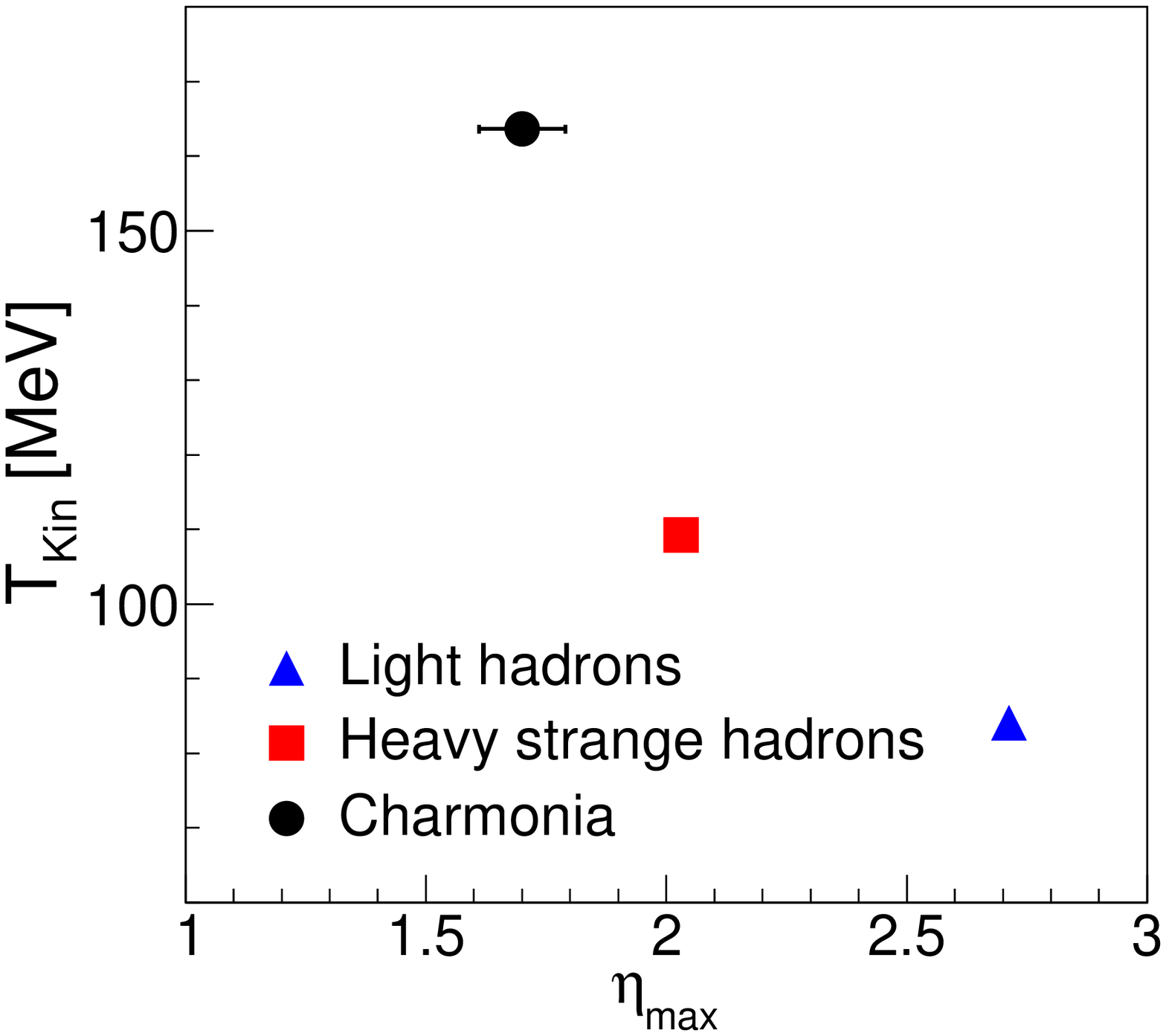}
\includegraphics[scale=0.28]{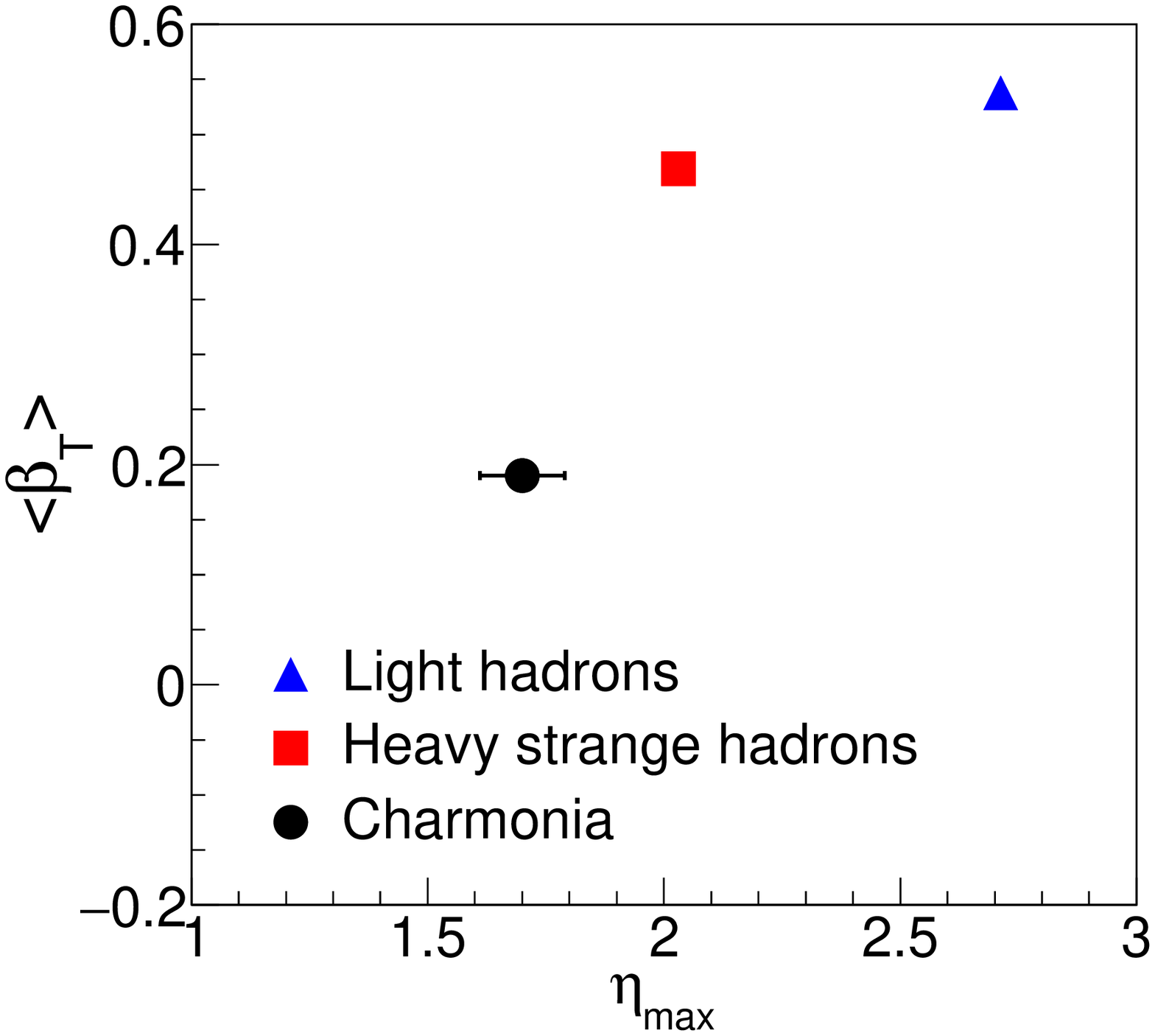}
\caption{The (partial) expansion history of the fireball created in central Pb+Pb collisions at $158A$ GeV. The points indicate the temperature ($T_{kin}$) and mean transverse collective flow velocity ($\langle \beta_{T} \rangle$) of the system at the time of charm kinetic freeze-out (filled circle), heavy strange kinetic freeze-out (filled square) and light hadron kinetic freeze-out (filled triangle). The error on $\eta_{max}$ for charmonia is assigned by varying it in such a way that the corresponding $\chi^{2}/N_{\rm dof}$ increases in magnitude by unity compared to it's minimum value. Errors on rest of the parameters are within the marker size.}
\label{158AGeV}
\end{figure*}
%

Details of the non boost-invariant blast wave model and its adopted version employed for the present analysis can be respectively found in Ref.~\cite{Dobler:1999ju} and Ref.~\cite{Rode:2018hlj}. Within this model, the thermal single particle spectrum for central collisions, in terms of transverse mass $m_T (\equiv\sqrt{p_T^2+m^2})$ and rapidity $y$ are given by
\begin{eqnarray}\label{therm}
   \frac{dN}{m_{T} dm_{T}  dy}
   &=&
   \frac{g}{2\pi}m_{T}\tau_F
   \int_{-\eta_{\max}}^{+\eta_{\max}} d\eta\,\cosh(y-\eta)
 \nonumber\\
   &\times&
   \int_0^{R(\eta)} r_\perp\,dr_\perp \, 
   \I_0\left(\frac{p_\T\sinh\rho(r_\perp)}{T}\right) \\
   &\times&
   \exp\left(\frac{\mu-m_\T\cosh(y{-}\eta)\cosh\rho(r_\perp)}
                  {T}\right) .
 \nonumber 
\end{eqnarray}
where $g$ denotes the degeneracy of particle species and $\eta\equiv\tanh^{-1}(z/t)$ is the space-time rapidity. In the transverse plane, the flow rapidity (or transverse rapidity) $\rho$ is related to the collective transverse fluid velocity, $\beta_{T}$, via the relation $\beta_{T}=\tanh(\rho)$. Assumption of the instantaneous common freeze-out of the fireball makes freeze-out time $\tau_F$ independent of the transverse co-ordinate $r_{\perp}$. In the spirit of a Hubble like expansion of the fireball in the transverse plane, the radial dependence of the transverse fluid velocity is assumed to be of the form: 
\begin{equation}
\label{beta}
\beta_{T}(r_{\perp}) = \beta^{0}_{T}\left( {\frac{r_{\perp}}{R}} \right). 
\end{equation}
where $\beta^{0}_{T}$ is the transverse fluid velocity at the surface of the fireball. Note that in the above equation, we have $R$ in the denominator as opposed to $R_0$ in the model of Ref.~\cite{Dobler:1999ju} where they parametrized the transverse rapidity. Due to this feature, the transverse flow vanishes at the center and assumes maximum value $\beta^{0}_{T}$ at the edges of the fireball as $r_{\perp}\rightarrow R$. For such a linear parametrization, the average transverse flow velocity is given by $\langle\beta_T\rangle=\frac{2}{3}\beta^{0}_{T}$ and therefore does not depend on $\eta$.

Considering reflection symmetry about the center of mass, the freeze-out volume is constrained in the region $-\eta_{max}\le\eta\le\eta_{max}$, to account for the limited available beam energy. In the transverse plane, the fireball is considered to have an elliptic shape, and the transverse size is parameterized as 
\begin{equation}\label{ellipsoid}
R(\eta) = R_0\,\sqrt{1 - {\eta^2\over\eta_{\max}^2}}\,,
\end{equation}
where $R_0$ is the transverse size of the fireball at $\eta=0$. Note that changing the integral variable $r_\perp\to r_\perp/R$ in Eq.~\eqref{therm}, the dependence on $R_0$ factors out leading to an overall volume factor $\tau_F R_0^2$. Moreover, dependence of system boundary in the transverse plane on the longitudinal coordinate, as parameterized in Eq.~\eqref{ellipsoid}, removes the assumption of boost-invariance. On the freeze-out surface, the temperature is taken to be constant and the transverse flow gradient along $r_{\perp}$ is constant with respect to $r_{\perp}$ and depends only on $\eta$ via $R(\eta)$. One can also note from Eq.~\eqref{therm} that the integral variable $r_{\perp}$ varies between $0 \le r_{\perp} \le R(\eta)$. Therefore, even though $R(\eta) \rightarrow 0$ as $\eta \rightarrow \pm{\eta_{max}}$, the transverse velocity $\beta_{T}(r_{\perp})$ given in Eq.~\eqref{beta} is always finite and lies in the physical range (preserves causality) provided $\beta^{0}_{T}<1$. From our analysis of the experimental results, we find that the extracted value of $\beta^{0}_{T}$ indeed  never leads to causality violation. From Eq.~\eqref{beta}, we also see that the transverse flow gradient along $r_{\perp}$ diverges as $\eta \rightarrow \pm{\eta_{max}}$. This may potentially lead to failure of the model when one tries to incorporate quantities which depend on gradients, for instance dissipative effects. However, in the present framework we do not have such gradients as we are dealing with non-dissipative blast wave model and therefore one does not encounter any issues in the model implementation.

%

%


\section{Results and discussions}

%
\begin{figure}[ht]
\includegraphics[scale=0.4]{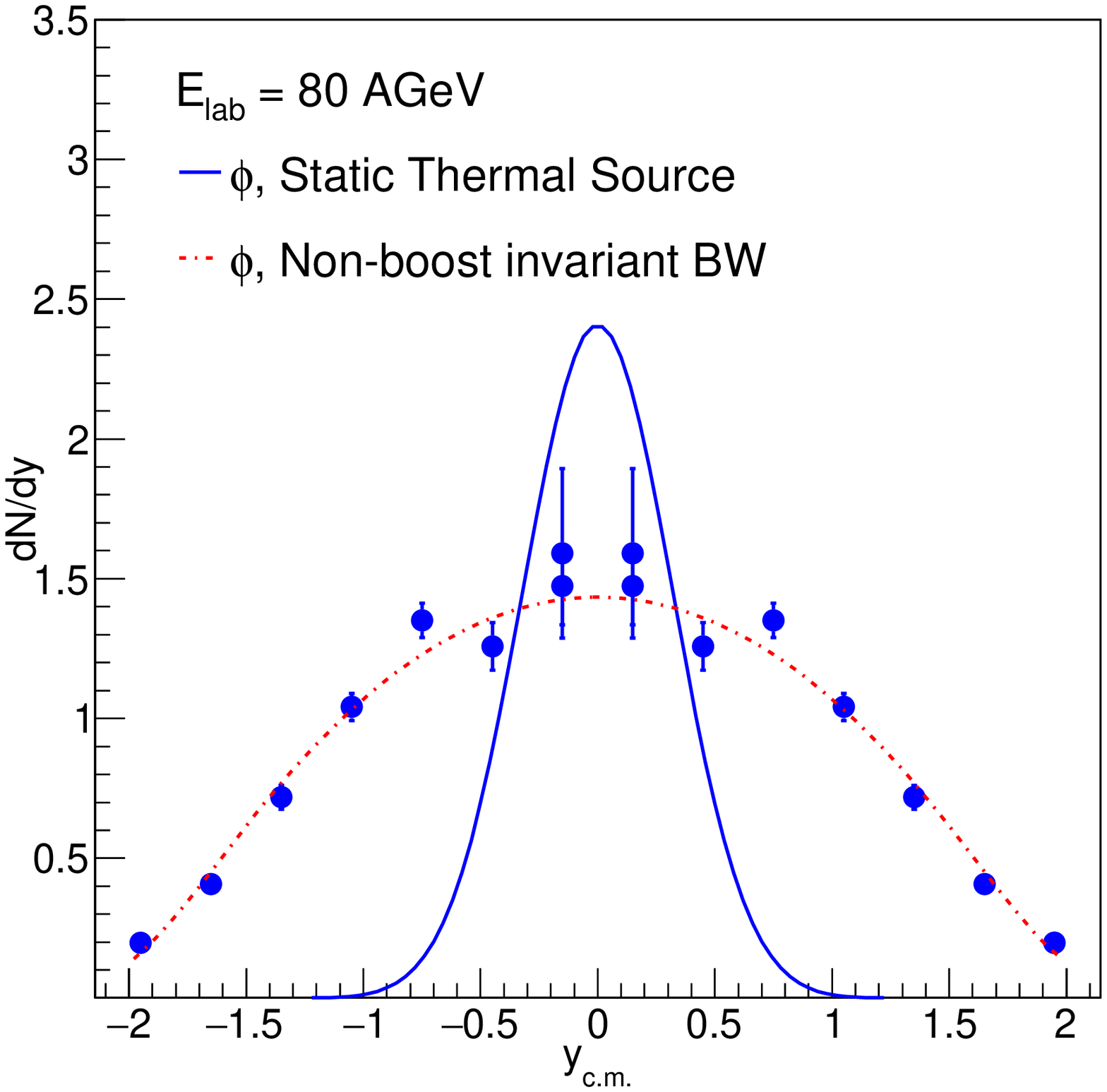}
\caption{Rapidity density distributions of $\phi$ meson in 80A GeV central Pb+Pb collisions at SPS. Data are compared with predictions from a static thermal model (simple continuous line) and non boost-invariant blast-wave model (dot dashed line). The source temperature in both cases is fixed to be $T_{kin} \simeq 106$ MeV. Vertical bars indicate the statistical errors.}
\label{compdiffrap}
\end{figure}
%

%
\begin{table*}[ht]\centering
\vglue4mm
\begin{tabular}{cccccc} \hline
$\rm E_{Lab}$ & Hadrons & squared sound  & $\chi^{2}/N_{\rm dof}$\\
(A GeV) &  &  velocity ($c^{2}_{s}$)  & \\ \hline
20 \rule{0pt}{0.5cm} &Heavy strange&$0.1602 \pm 0.0006$& 2.4\\
&Light&$0.0755 \pm 0.0000$& 331.7\\\hline
30 \rule{0pt}{0.5cm} &Heavy strange&$0.2156 \pm 0.0009$  & 2.2\\
&Light&$0.121 \pm 0.0000$  & 238.7\\\hline
40 \rule{0pt}{0.5cm} &Heavy strange& $0.2215 \pm 0.0007$  & 2.1\\
&Light& $0.1682 \pm 0.0001$  & 38.4\\
\hline
80 \rule{0pt}{0.5cm} &Heavy strange&$0.2234 \pm 0.0005$  & 2.9\\
&Light& $0.2136 \pm 0.0000$  &22.2\\
\hline
158 \rule{0pt}{0.5cm} &Heavy strange&$0.2511 \pm 0.0003$  & 3.1\\
&Light& $0.2276 \pm 0.0001$  & 26.5\\
\hline
\end{tabular}
\caption{Summary of the fit results (squared speed of sound ($c^{2}_{s}$) and $\chi^{2}/N_{\rm dof}$ values) of Rapidity spectra of heavy strange and light hadrons at different energies from SPS using non-conformal solution of the Landau hydrodynamics.}
\label{tabIII}
\end{table*}
%

The results of our analysis have been presented in this section. The measured $p_{T}$ and $y$ spectra of all the available heavy strange hadrons produced in central Pb+Pb collisions from NA49 Collaboration~\cite{Alt:2008qm, Alt:2008iv, Alt:2004kq} at SPS, in the beam energy range $\rm E_{Lab}=20A-158A $ GeV, are analysed for this purpose. Not much data on strange hadrons are available in Au+Au collisions at AGS energies, except the measurements of $\Lambda$~\cite{Barrette:2000cb} and $\phi$~\cite{Back:2003rw} at 11.5 and 11.7 AGeV and with different kinematic coverage, from E877 and E917 experiments respectively. Nonetheless, we confine ourselves only to the SPS energy domain. Data on $p_T$ distribution of a variety of strange hadron species from STAR Collaboration~\cite{Adam:2019koz} at RHIC beam energy scan (BES) program are preliminary at the moment~\cite{Helen} and we have not included in the present analysis. Moreover, the corresponding $y$ distributions have also not been reported yet. The analysis of the data above SPS energies is beyond the scope of this work.

Details of the data sets of heavy strange hadrons under investigation are summarized in Table~\ref{tabI}. The lightest hadron in our chosen set is thus $\phi$ meson, having a mass of $1.02$ GeV. Therefore contributions from hadronic resonance decays are expected to be small and hence ignored. One may note that at all the selected collision energies, the $p_{T}$ spectra of $\phi$ mesons are available over rather wide rapidity bins. For uniformity, the $p_{T}$ spectra of all investigated species are evaluated at $y_{c.m.}=0$. Our fit results  remain almost unchanged if instead the transverse yields are estimated by integrating over the available particle rapidity window.
The model fits are done by minimizing the value $\chi^{2}/N_{dof}$, where $N_{dof}$ denotes the number of degrees of freedom that is the number of data points minus the number of fitting parameters. The MINUIT~\cite{minuit2} package as available in ROOT framework~\cite{root}, is employed for the minimization procedure in our analysis. 

For light hadrons, we have repeated the analysis of Ref.~\cite{Rode:2018hlj} for simultaneous fits to only $\pi^{-}$ and $K^{\pm}$ but using an updated iterative procedure which is discussed below. In the present work, we do not consider protons in the fits as the rapidity spectra of protons are not available at SPS energies. Moreover, due to stopping at low energies, all observed protons may not be thermally produced. However, we have checked that the main message of the present work remains unaltered irrespective of whether we include proton in light hadron or heavy hadron set. At $158$A GeV, data are also available for $p_T$-spectra of $J/\Psi$ and $\Psi'$ for which we perform a simultaneous fit as separate set.

%
\begin{figure}[t]
\includegraphics[scale=0.4]{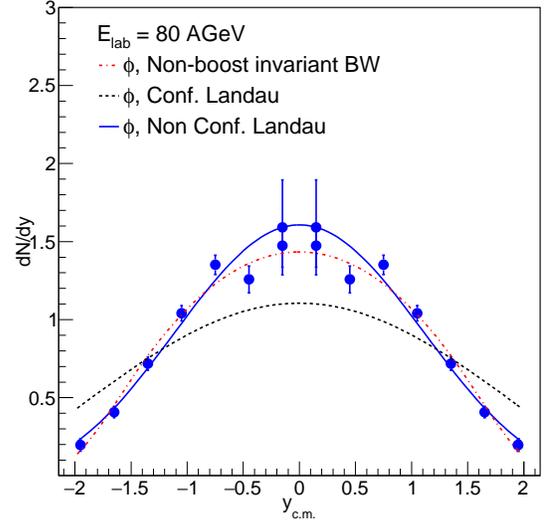}
\caption{Rapidity density distributions of $\phi$ meson in 80A GeV central Pb+Pb collisions at SPS. Data are compared with predictions from different dynamical models namely, non boost-invariant blast-wave model (dot dashed line), conformal (dotted line) and non-conformal (simple continuous line) solution of the Landau hydrodynamics. Statistical errors are shown as vertical bars.}
\label{compdiffrap2}
\end{figure}
%

%
\begin{figure*}[ht]
\begin{picture}(160,140)
\put(0,0){\includegraphics[scale=0.28]{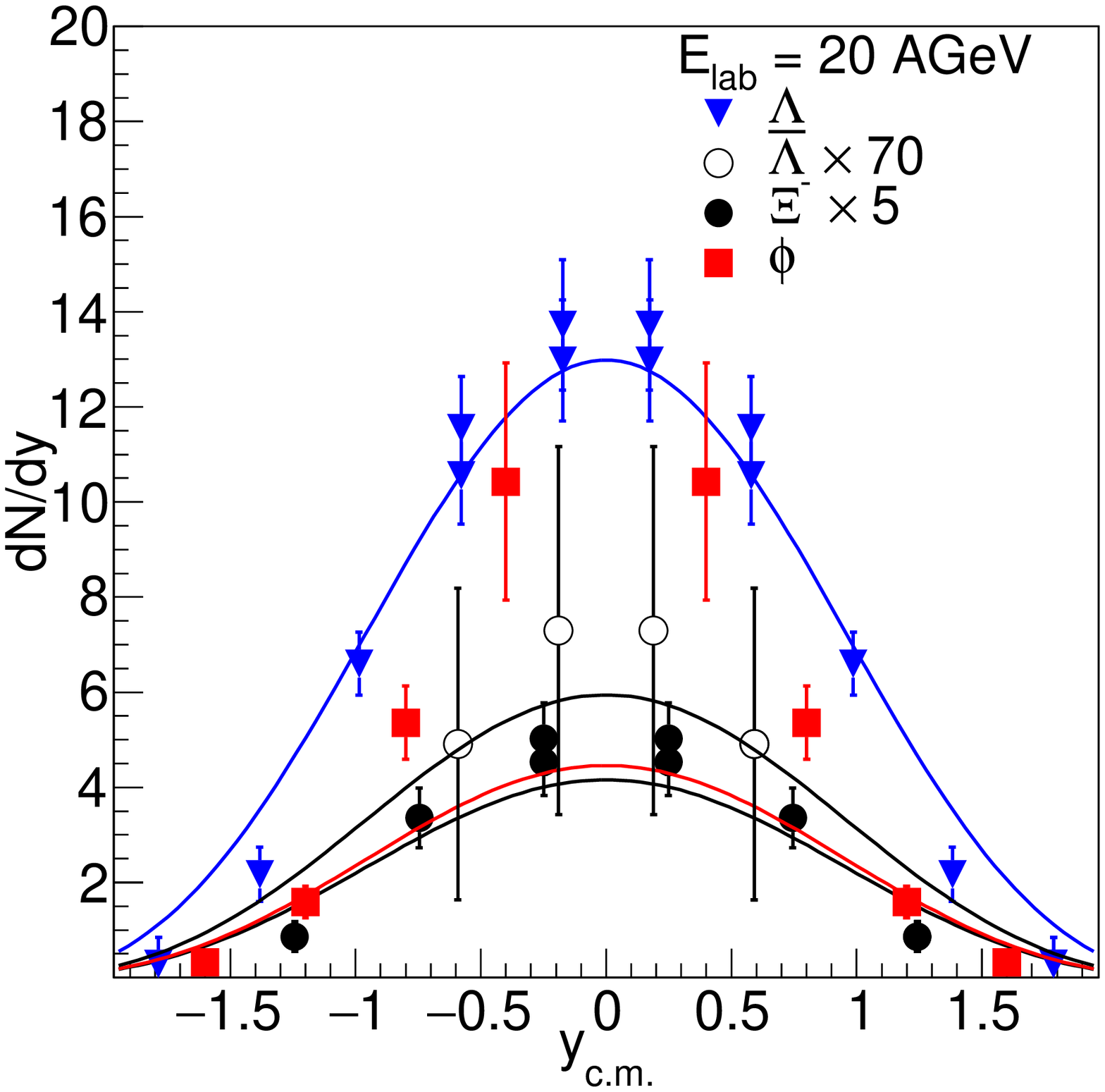}}
\put(40,110){(a)}
\end{picture}
\begin{picture}(160,140)
\put(0,0){\includegraphics[scale=0.28]{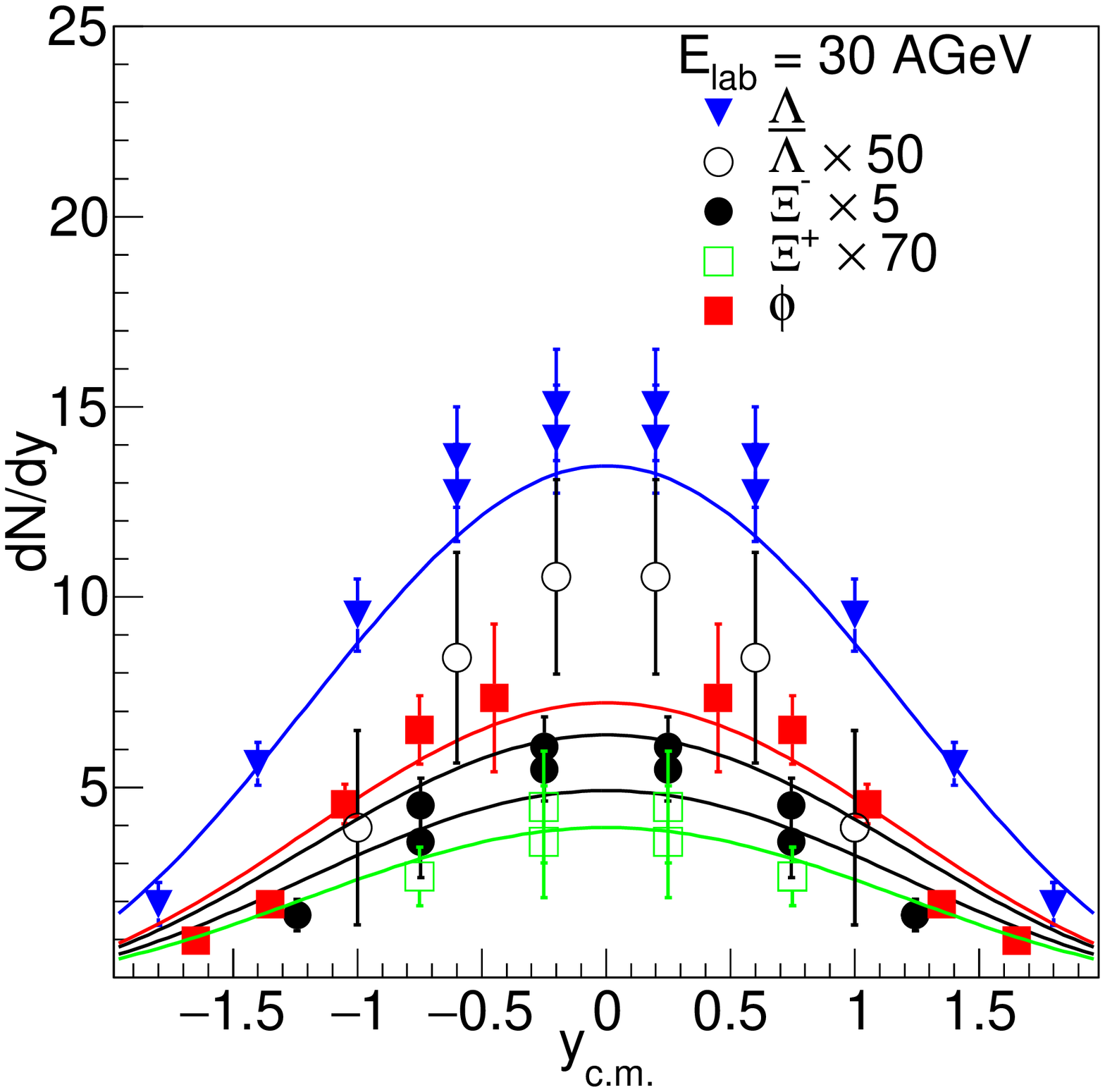}}
\put(40,110){(b)}
\end{picture}
\begin{picture}(160,160)
\put(0,0){\includegraphics[scale=0.28]{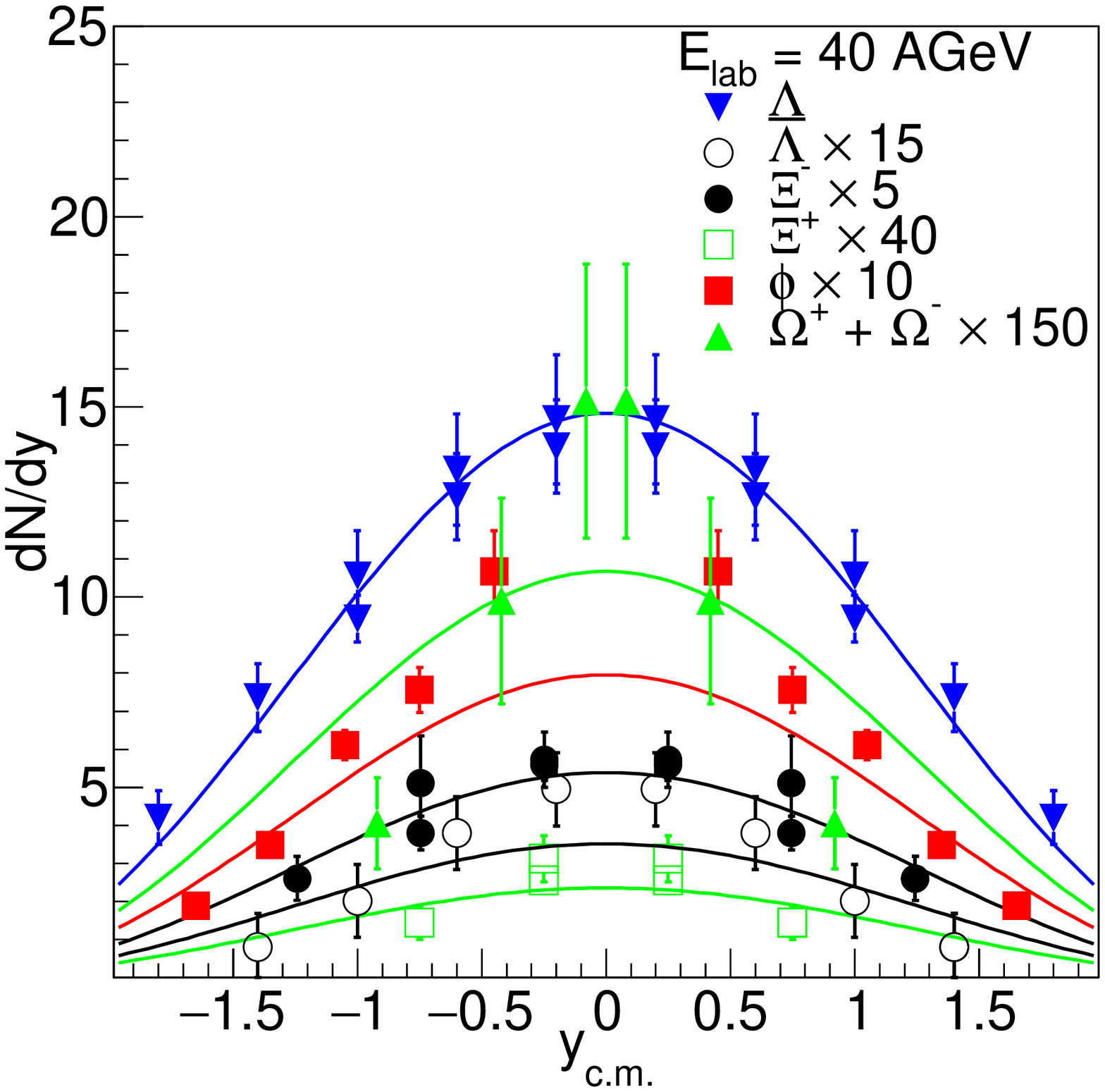}}
\put(40,110){(c)}
\end{picture}
\begin{picture}(160,160)
\put(0,0){\includegraphics[scale=0.28]{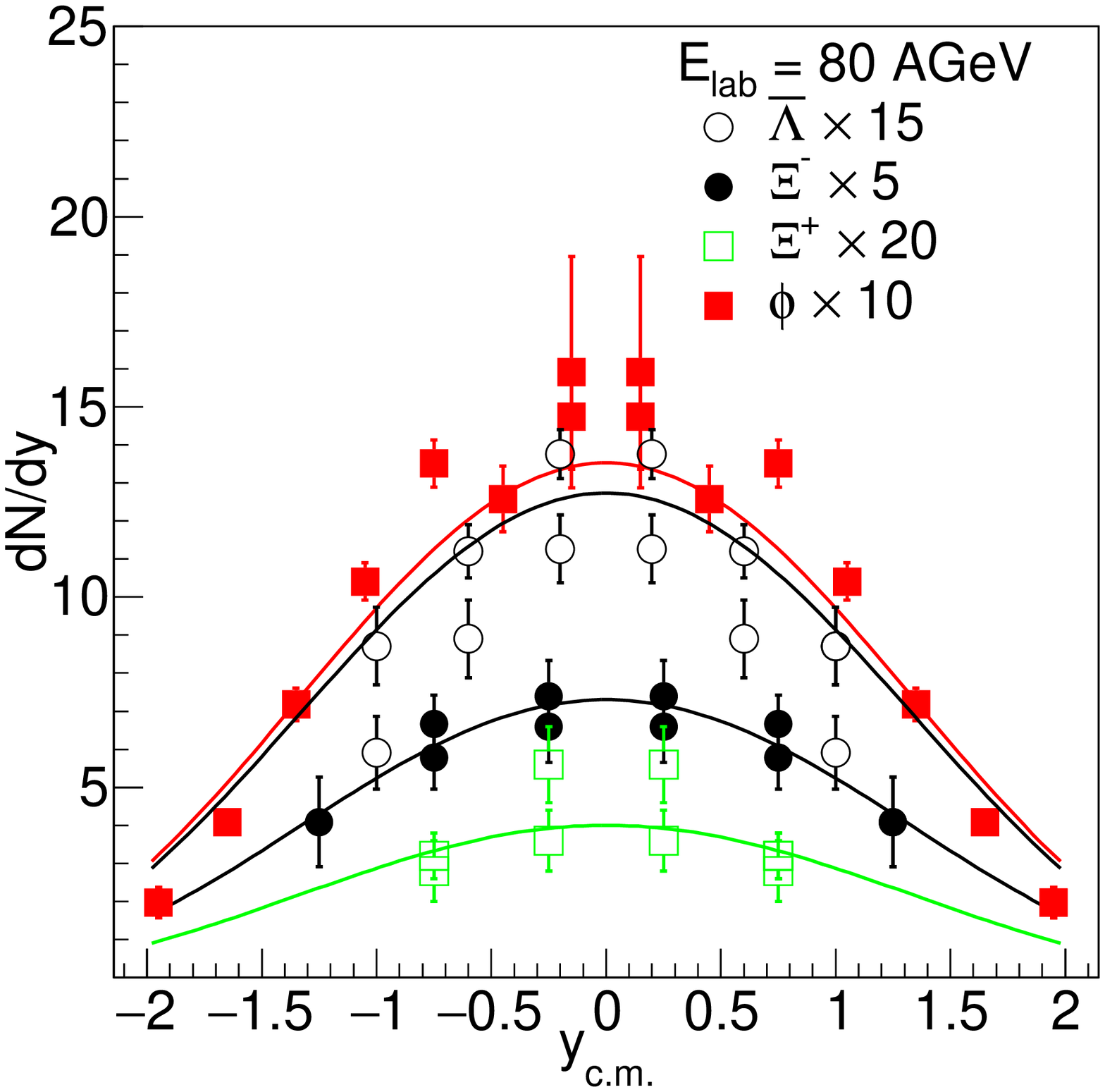}}
\put(40,110){(d)}
\end{picture}
\begin{picture}(160,160)
\put(0,0){\includegraphics[scale=0.28]{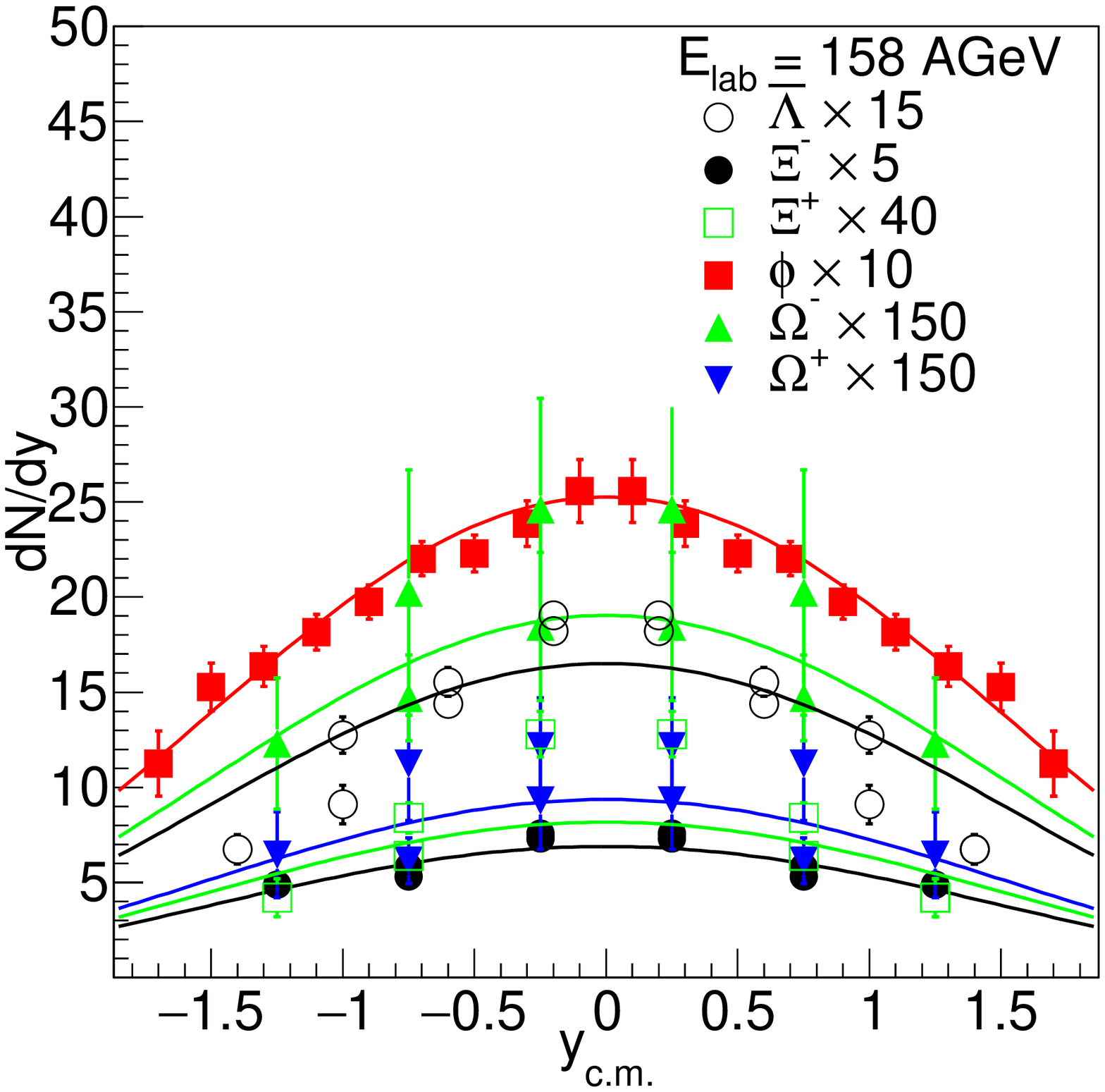}}
\put(40,110){(e)}
\end{picture}
\caption{Fitted rapidity distribution of $\Lambda$, $\bar{\Lambda}$, $\phi$, $\Xi^{\pm}$ and $\Omega^{\pm}$ using non-conformal solution of the Landau hydrodynamics in central Pb+Pb collisions from SPS at (a) 20A GeV, (b) 30A GeV, (c) 40A GeV, (d) 80A GeV and  (e) 158A GeV beam energies. Error bars indicate available statistical error.}
\label{fig11}
\end{figure*}
%

%
\begin{figure*}[t]
\begin{picture}(160,140)
\put(0,0){\includegraphics[scale=0.28]{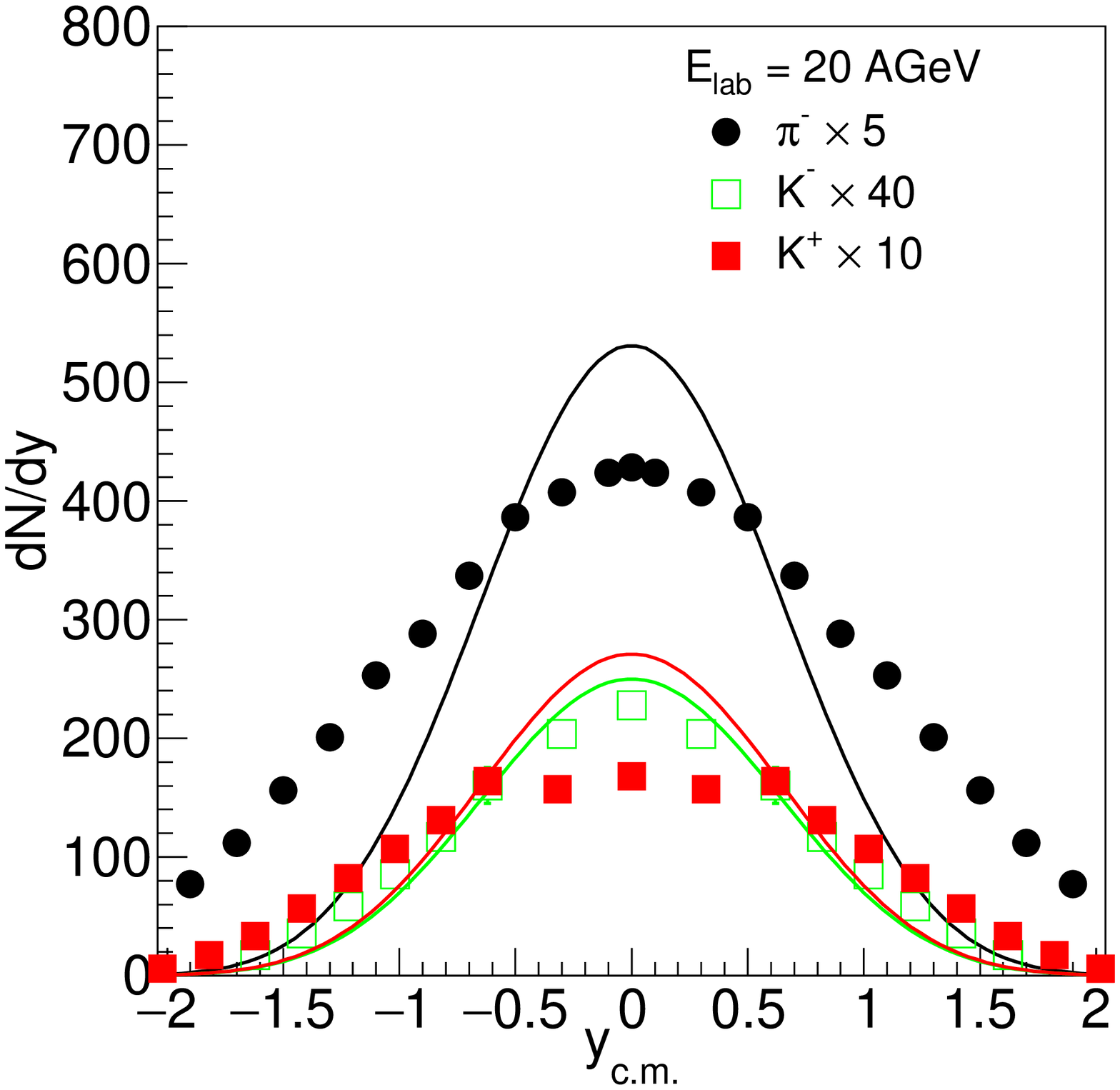}}
\put(40,110){(a)}
\end{picture}
\begin{picture}(160,140)
\put(0,0){\includegraphics[scale=0.28]{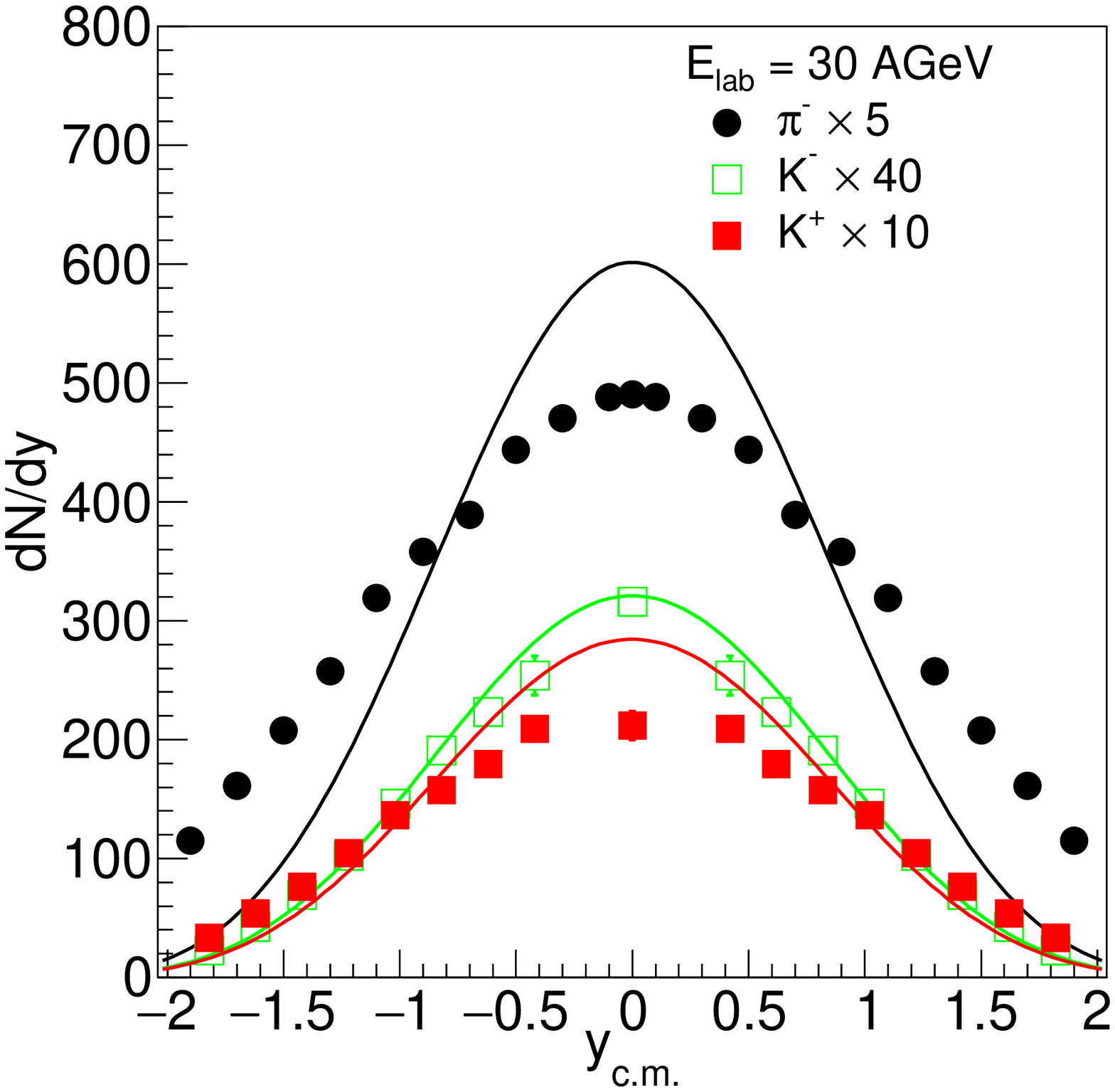}}
\put(40,110){(b)}
\end{picture}
\begin{picture}(160,160)
\put(0,0){\includegraphics[scale=0.28]{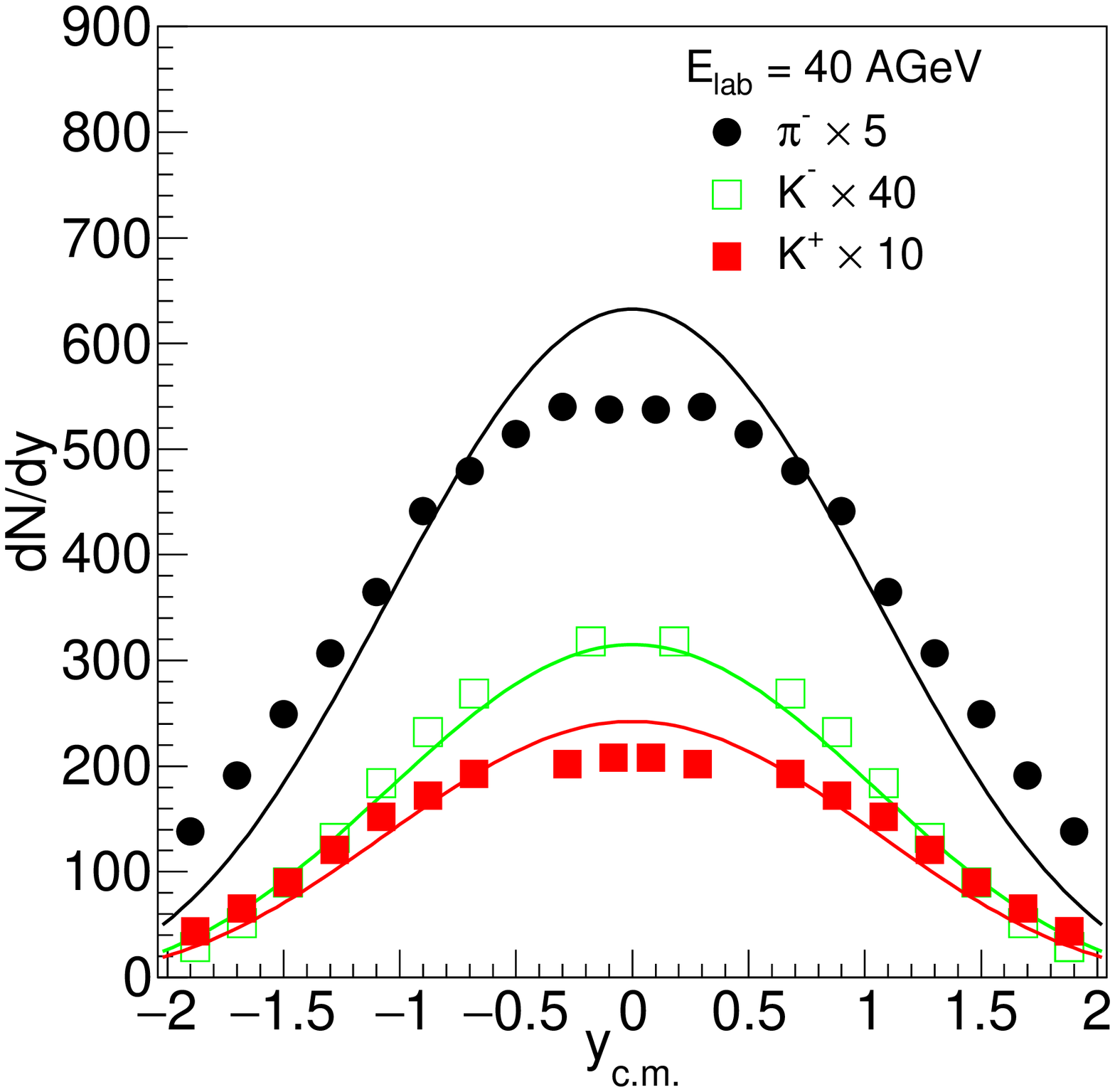}}
\put(40,110){(c)}
\end{picture}
\begin{picture}(160,160)
\put(0,0){\includegraphics[scale=0.28]{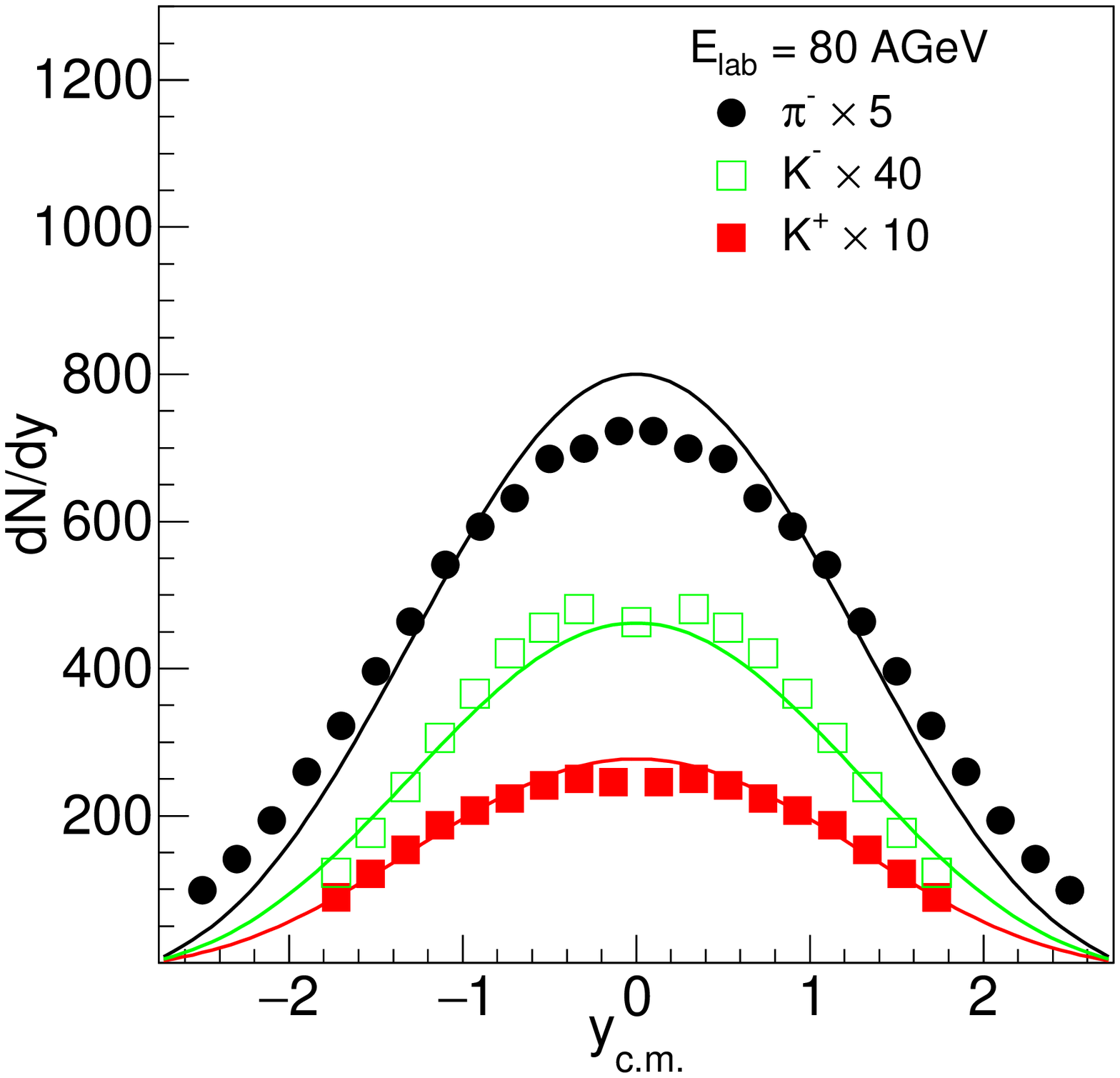}}
\put(40,110){(d)}
\end{picture}
\begin{picture}(160,160)
\put(0,0){\includegraphics[scale=0.28]{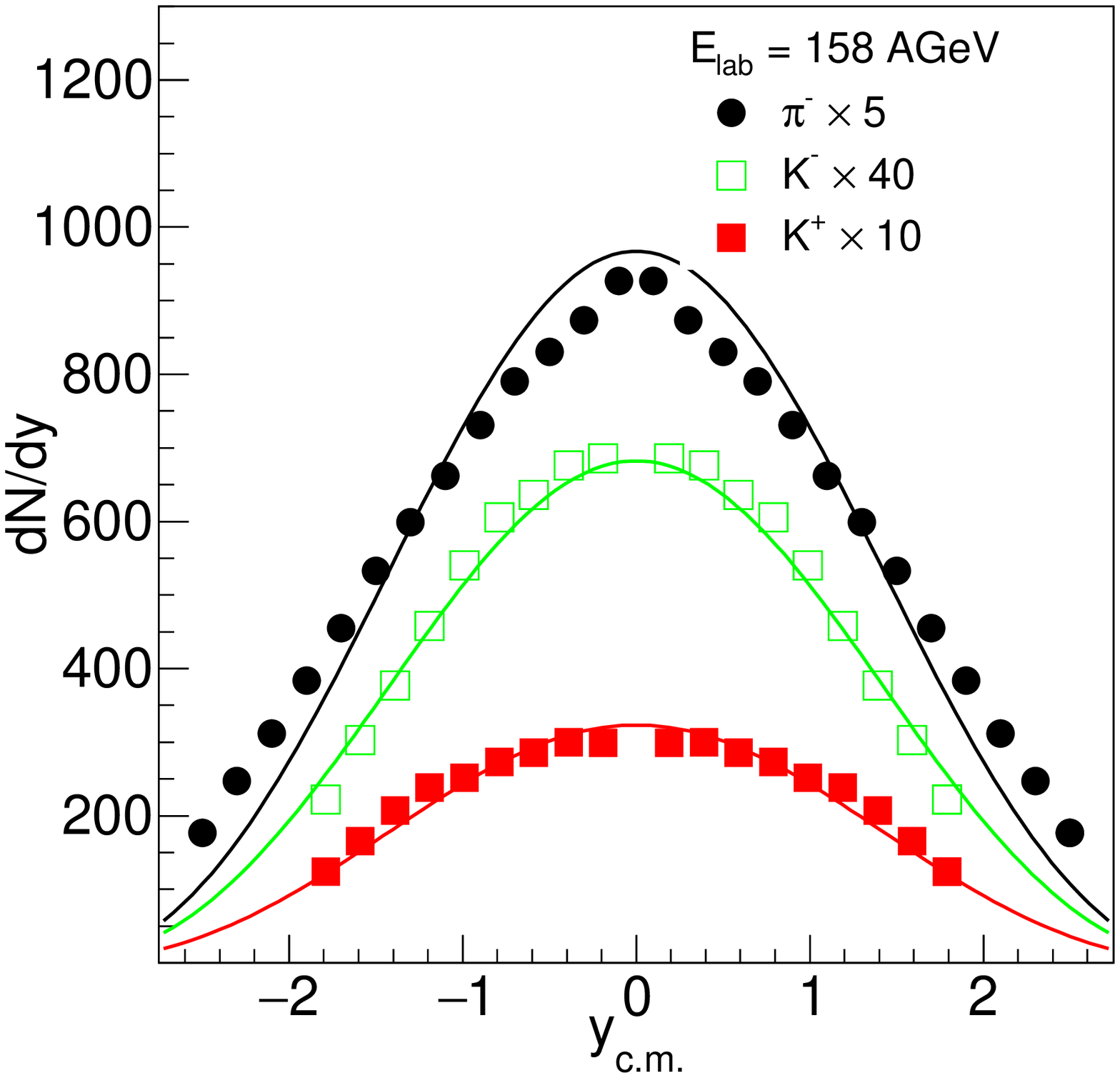}}
\put(40,110){(e)}
\end{picture}
\caption{Fitted rapidity distribution of $\pi^{-}$ and $\rm K^{\pm}$ using non-conformal solution of the Landau hydrodynamics in central Pb+Pb collisions from SPS at (a) 20A GeV, (b) 30A GeV, (c) 40A GeV, (d) 80A GeV and  (e) 158A GeV beam energies. Error bars indicate available statistical error.}
\label{fig13}
\end{figure*}
%

We start with analysis of $p_{T}$ spectra of the heavy strange hadrons. At a given collision energy, the $p_{T}$ distribution of all the available heavy strange hadrons are fitted simultaneously using Eq.~(\ref{therm}). To minimize the free parameters in the fit, the freeze-out time $\tau_F$, degeneracy factor $g$ and the fugacity (chemical potential) are coupled together into a single normalization constant $Z\equiv{\frac{g}{2\pi}}\tau_F\exp(\mu/T)$, which is adjusted separately for each particle species. Since chemical freeze-out fixes the value of chemical potential, its absorption inside the normalization would not affect the thermodynamic conditions at kinetic freeze-out. As mentioned earlier, the dependence on $R_0$ factors out leading to a volume factor $\tau_F R_0^2$ which can also be absorbed inside the overall normalization. Thus we are left with three free parameters, namely $T_{kin}$, $\eta_{max}$ and $\beta^{0}_{T}$.

It is important to note that, out of the three parameters, $T_{kin}$, $\eta_{max}$ and $\beta^{0}_{T}$, the rapidity spectra is more sensitive to the width in space-time rapidity $\eta_{max}$ and is not affected significantly by small changes in the other two parameters. On the other hand, the $p_T$ spectra is more sensitive to $T_{kin}$ and $\beta^{0}_{T}$ and small changes in $\eta_{max}$ does not affect the slope of the $p_T$ spectra. Here we adopt an iterative procedure to obtain the best fit values of the parameters. At a given collision energy, the value $\eta_{max}$ is first fixed from the simulatenous fits of the rapidity distributions of the available heavy strange hadrons with some initial guess values of $T_{kin}$ and $\beta^{0}_{T}$. Subsequently, we use this $\eta_{max}$ to fit the corresponding $p_{T}$ distributions and extract the values of $T_{kin}$ and $\beta^{0}_{T}$. These extracted values of $T_{kin}$ and $\beta^{0}_{T}$ are now used to fit the rapidity spectra again to obtain a refined value of $\eta_{max}$. This iterative procedure converges quickly and we obtain the fitted values to desired accuracy. Simultaneous fits of the $p_{T}$ and the rapidity spectra, as described above, at 20A, 30A, 40A, 80A and 158A GeV are performed for all available heavy strange hadrons. The best fit results for the $p_{T}$ and rapidity spectra are shown in Fig.~\ref{fig4} and \ref{fig7} respectively. Note that we refrain from fitting the $\Lambda$ rapidity distributions at 80A and 158A GeV because of incomplete stopping at these energies and the fact that $\Lambda$  carry significant fraction of total net baryon number, its rapidity distributions are flat~\cite{Anticic:2003ux, Alt:2008qm}. The fit to the data is  well described by the single $\eta_{max}$, $\langle \beta_{T} \rangle$ and $T_{kin}$ values as can be observed from the $\chi^{2}/N_{\rm dof}$ values given in Table \ref{tabII}.

In Fig.~\ref{contour}, we show the two dimensional projection of the $\chi^{2}$ contour plots in the $\langle\beta_T\rangle - T_{\rm kin}$ plane for (a) 20A GeV, (b) 30A GeV, (c) 40A GeV, (d) 80A GeV and (e) 158A GeV beam energies. Different colors correspond to different values of $\chi^{2}/N_{\rm dof}$ and signify the uncertainty region around the best fit parameters. We see that the two parameters are anti-correlated, in the sense that a slight reduction in $T_{\rm kin}$ can be compensated by a concomitant enhancement in $\langle\beta_T\rangle$ with a very small effect on the quality of the fit to $p_T$ spectra. Driven by the associated experimental errors, which are purely statistical in nature, the uncertainty of the best fit parameters is larger at lower beam energy.

In Fig.~\ref{fig9}, we plot the extracted best fit parameters for heavy strange hadrons, namely, average transverse velocity ($\langle\beta_{T}\rangle$) and kinetic freeze-out temperature ($T_{kin}$) and $\eta_{max}$ as a function of the beam energy ($E_{Lab}$) and compare them with the corresponding values for light hadrons ($\pi^{-}$, $K^{\pm}$). However in our earlier work in Ref.~\cite{Rode:2018hlj}, the iterative fitting procedure described above was not followed. For consistency, we have refitted the pion and kaon spectra following the present iterative strategy and used the results of this improved fit in the current analysis. Note that the best fit values of the extracted parameters are found to be nearly unaltered. The $\eta_{max}$ value changes within 2--3 $\%$ compared to the older value, whereas changes in $T_{kin}$ and $\langle\beta_{T}\rangle$ are negligible. All the three quantities show increasing trend as a function of beam energy ($E_{Lab}$). Moreover, at all collision energies, the extracted temperatures are larger than those for light hadrons. Also the corresponding smaller $\langle \beta_{T} \rangle$ and $\eta_{max}$ values indicate the heavy strange particles decouple from the fireball earlier in time compared to the light hadrons. Thus the kinetic freeze-out also seem to exhibit a hierarchical structure, with more massive particles leaving the medium earlier in time.

As mentioned earlier, in Ref.~\cite{Gorenstein}, the $m_{T}$ spectra of J/$\psi$, $\psi'$  and $\Omega$ produced in $158A$ GeV central Pb+Pb collisions were analyzed within boost-invariant blast-wave dynamics. Based on the hypothesis that these heavy hadrons are produced via statistical coalescence and undergo freeze-out during hadronization, due to their small rescattering cross-sections in hadronic phase, an average transverse collective flow velocity of $\langle \beta_{T} \rangle \simeq 0.2$ was extracted from simultaneous fit to the spectra, restricting $T_{kin} = 170$ MeV, from analysis of hadron multiplicities.

For us it would be worth analyzing the available transverse distribution of charmonia in $158A$ GeV Pb+Pb collisions, measured by NA50 Collaboration~\cite{Ale05}, within the present model framework. Instead of fixing $T_{kin}$, we keep it free with other two parameters. The unavailability of the rapidity spectra dictates us to fix the three parameters from the $p_{T}$ spectra alone. Simultaneous fitting of J/$\psi$ and $\psi'$~\cite{Abreu:2000xe} $p_{T}$ distributions in the rapidity range (0 $\leqslant$ $y_{c.m.}$ $\leqslant$ 1) shown in Fig.~\ref{JPsi158AGeVpt}, gives the following values of the parameters: $T_{kin} = 164$ MeV, $\eta_{max} = 1.70$ and $\langle\beta_{T}\rangle \approx 0.2$, indicating the emission of these heavy resonances from the fireball much earlier in time. 

In absence of the rapidity spectra, the precision of the $\eta_{max}$ value for charmonia, extracted from $p_T$ spectra might be questionable. To decide the associated uncertainty in $\eta_{max}$, we adopt the following strategy. The value of $\eta_{max}$ is varied around the obtained value while keeping the other two parameters fixed to their respective best fit values, in such a way that the resulting $\chi^{2}/N_{\rm dof}$ increases in magnitude by unity from it's minimum value. The corresponding variation in $\eta_{max}$ is assigned as the error on the parameter. Moreover, we have considered the entire $p_T$ range of charmonia available in the data for the analysis. 

One might question the applicability of boosted thermal models to describe $p_T$ values as high as $5$ GeV, in order to remain in the region sensitive to collective effects and free from hard scattering processes. The fit parameters remain essentially the same if we limit the fitted $p_T$ range up to much lower value, say $2$ GeV. Since the rest masses of $J/\psi$ and $\psi'$ mesons are of the order of 3 GeV, they remain thermalized to much higher $p_T$ compared to the light hadrons. The NA60 Collaboration~\cite{Arn07} has measured $J/\psi$ production in $158A$ GeV In+In collisions. However, the corresponding transverse distributions have not been published yet. Note that we exclude $\Omega$  baryon, as it is a member of our heavy strange set at $80A$ and $158A$ GeV and much lighter than the charmonium family.

In Fig.~\ref{158AGeV}, we show the freeze-out points extracted from the measured transverse spectra of hidden charm, heavy strange and light hadrons at 158A GeV, defining the path of the expanding system in the $T_{kin}$-$\langle \beta_{T} \rangle$ plane (left panel), $T_{kin}$-$\eta_{max}$ plane (center panel) and $\langle \beta_{T} \rangle$-$\eta_{max}$ plane (right panel). Results show a monotonous behavior which support a clear existence of a mass dependent hierarchy in thermal freeze-out of hadrons. This hierarchy of kinetic freeze-out is expected as the medium induced momentum change of heavy hadrons would be smaller compared to lighter hadrons. Hence, as the temperature of the fireball decreases, one would expect an earlier kinetic decoupling of heavy hadrons. Therefore, with a systematic investigation of the freeze-out parameters of different hadron species one can in principle trace (partially) the expansion history of the fireball produced in nuclear collisions. Till date no charm data are available in heavy-ion collisions below top SPS energy. The upcoming NA60$+$ experiment at SPS~\cite{Agnello:2018evr}, aims at the measurement of charmonia in $20A - 158A$ GeV Pb+Pb collisions. Data once available at lower energies will be enable to concretely establish this mass dependent hierarchy in thermal freeze-out.

At this juncture, it would be worth investigating the robustness of the extracted parameters against the small changes in the model inputs. The linear flow profile, that we have chosen seems to be quite reasonable for central collisions and validated by hydrodynamical calculations~\cite{Teaney:2001av}. However there are alternative formulations of blast wave models, where the flow profile ($n$) is set as free parameter and fixed from the data. In a recent work~\cite{Melo:2019mpn}, the kinetic freeze-out conditions in central heavy-ion collisions are investigated over a wide energy range from $\sqrt{s_{NN}} = 7.7$ GeV to $\sqrt{s_{NN}} = 2.76$ TeV. The single-particle $p_T$ spectra of identified bulk hadrons ($\pi$, $K$, $p$) are fitted using a boost-invariant blast-wave framework. In addition to $T_{kin}$ and $\langle\beta_{T}\rangle$, $n$ is also fixed from fit to $p_T$ spectra. The transverse flow profile is found to be least sensitive parameter and roughly consistent with a constant value $n \simeq 0.75$ provided the fitted $p_T$ spectra at lower energies are limited to short range. 

In our case, if we keep $n$ as free parameter to fit the $p_T$ spectra, the fitted values of the exponent $n$ lies in the range of $1-1.15$, without any significant change in the rest of the fit parameters. On the other hand the opted parametrization of $R(\eta)$ describing an ellipse in the $\eta - r_{\perp}$ space is driven by the loss of cylindrical symmetry in low energy collisions~\cite{Nix} and the choice explicitly breaks the longitudinal boost-invariance. A wide range of possibilities were investigated in Ref.~\cite{Dobler:1999ju}: namely cylindrical vs. elliptic transverse geometry; constant vs. $\eta$-dependent freeze out temperature and a constant vs. $\eta$-dependent transverse flow gradient. Combining elliptic geometry with constant temperature and transverse flow gradient appeared to be an optimum choice giving a reasonably good description of both $p_T$ and $y$ distributions. A further fine-tuning by combining the elliptic geometry with an additional $\eta$-dependence of freeze-out temperature leads to significant increment in the computing time but minor improvement in results. We thus desist from using any alternative parameterizations for transverse shape, freeze-out temperature and transverse flow gradient in our analysis.

Before we move forward, it might be interesting to note that the possible existence of hierarchy in the kinetic freeze-out parameters of the produced particles has been studied earlier at RHIC and LHC energies. In Ref.~\cite{Chatterjee:2014lfa}, the authors have analyzed the $p_T$ spectra of the identified hadrons in $\sqrt{s_{NN}} = 2.76$ TeV Pb+Pb collisions, using a so-called longitudinal boost-invariant single freeze-out model, which describe both the particle spectra and particle ratios with a single value of the temperature. Their results indicated a flavour dependent kinetic freeze-out scenario, with strange hadrons leaving the fireball earlier in time than the non-strange hadrons.  In Refs.~\cite{Thakur:2016boy,Lao:2015zgd,Khuntia:2018znt}, the authors have also analyzed the $p_T$ spectra of different particle species measured at mid-rapidity in p+p and A+A collisions at various collision energies at RHIC and LHC, using different variants of Tsallis distribution. The freeze-out temperature is found to increase with the increase in particle mass, exhibiting an evidence of mass dependent multiple kinetic freeze-out scenario. In fact the dependence of the inverse slope parameter of the $p_{T}$ spectra (effective temperature) of the identified hadrons emitted in central heavy-ion collisions at 158 A GeV Pb+Pb collisions at SPS and $\sqrt{s_{NN}} = 200$ GeV Au+Au collisions at RHIC were first reported in so the so-called ``Nu-Xu''~\cite{vanHecke:1999jh} plot, representing the freeze-out systematics for a set of hadronic species.

We also investigate the effect of longitudinal flow on the observed rapidity distribution of the heavy strange hadrons. We have seen earlier~\cite{Rode:2018hlj} that isotropic emission from static thermal model cannot describe the measured rapidity distribution of light hadrons at all beam energies. Collective expansion in the longitudinal direction is essential to reproduce the data. An illustrative comparison to understand how the longitudinal motion influence the rapidity distribution of the heavy strange hadrons is presented in Fig.~\ref{compdiffrap}. The rapidity distribution of $\phi$ mesons measured in $80A$ GeV central Pb+Pb collisions is contrasted with that from a static thermal model as well as from the present blast-wave model calculations. For both the cases the source temperature is fixed to $T_{kin} \simeq 106$ MeV. The rapidity distribution as obtained from static isotropic thermal source falls much faster than the data, a feature that is common for all heavy strange hadrons and at all investigated energies. The feature also holds true if one attempts to fit the rapidity spectra by static isotropic thermal source with temperature kept as free parameter.

This essentially completes our study of kinetic freeze-out conditions for heavy strange hadrons within non boost-invariant blast wave model. However, before we close, it might be useful to take a deeper look at the longitudinal dynamics particularly so due to the absence of boost-invariance at low energy collisions. Hence moving forward, the longitudinal properties of the medium are further explored by fitting rapidity spectra of heavy strange hadrons at beam energies 20A, 30A, 40A, 80A and 158A GeV using a different prescription, available in literature. 

The rapidity distribution as predicted by the recently developed non-conformal solution of the Landau hydrodynamics is given by~\cite{Biswas:2019wtp},
\begin{align} \label{dndynonconflandau}
\dfrac{dN}{dy} \sim \exp\left[\frac{1-c_s^2}{2c_s^2} \sqrt{{y_b'}^2-y^2} \right],
\end{align}
where, $c_s^2$ is the squared sound velocity in the medium, $y_b'\equiv\frac{1}{2}\ln[(1+c_s^2)/(4c_s^2)]+y_b$, with $y_b=\ln[\sqrt{s_{NN}}/m_p]$ being the beam rapidity and $m_p$ the proton mass. The conformal solution of Landau hydrodynamics can be restored by putting $c_s^2=1/3$ \cite{Wong:2008ex}. In Fig.~\ref{compdiffrap2}, we compare the available data on rapidity distribution of $\phi$ mesons in 80A GeV central Pb+Pb collisions at SPS, with predictions from different dynamical models. We find that the rapidity spectra from conformal solution falls off too slowly and does not give good agreement with the data. On the other hand, both blast-wave as well as the non-conformal solution of Landau hydrodynamics explain the data really well.

%
\begin{figure}[t!]
\includegraphics[scale=0.42]{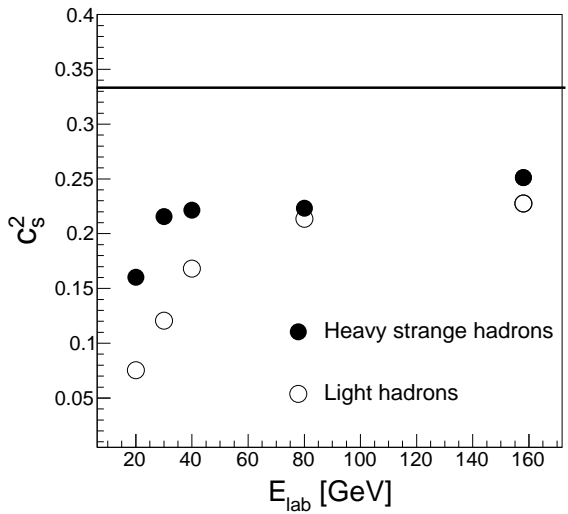}
\caption{Variation of the speed of sound for heavy strange and light hadrons with beam energy, extracted from fit to the longitudinal spectra using non-conformal solution of Landau hydrodynamics. Horizontal line at $\rm c^{2}_{s}=1/3$ indicate the ideal gas limit. Errors are within the marker size.}
\label{fig15}
\end{figure}
%

%
\begin{figure}[ht]
\includegraphics[scale=0.42]{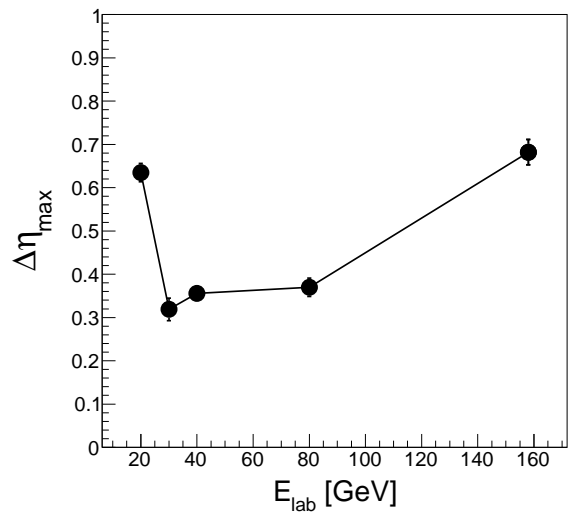}
\caption{Difference of $\eta_{\rm max}$ for light hadron and heavy strange hadrons, $\Delta\eta_{\rm max}=|\eta_{\rm max}^{\rm LH}-\eta_{\rm max}^{\rm HS}|$ as a function of beam energy.}
\label{Delta_LH_HS}
\end{figure}

%

This motivates us to perform simultaneous fits to the available rapidity spectra of heavy strange hadrons using the non-conformal solution of the Landau hydrodynamics given in Eq.~\eqref{dndynonconflandau}. We obtain reasonably good fits, as shown in Fig.~\ref{fig11}, with good $\chi^{2}/N_{\rm dof}$ and the extracted values of $c_s^2$ are shown in Table~\ref{tabIII}. Here, $c_s^2$ is a common parameter for all species and only the overall normalization constant is allowed to be different. In practice the sound velocity, $c_s$, depends on temperature and thus varies during evolution of the expanding medium formed in heavy-ion collisions. However the analytical expression obtained in Ref.~\cite{Biswas:2019wtp} assumes constant value of $c_s^2$ and therefore our extracted $c_s^2$ correspond to an effective mean value. We have also used Eq.~\eqref{dndynonconflandau} to simultaneously fit the rapidity spectra of light hadrons at SPS energies as shown in Fig.~\ref{fig13}. However the model does not seem to work well for light hadrons as evident from the rather poor fit quality and associated huge values of $\chi^{2}/N_{\rm dof}$. Resultant values of the $c_s^2$ are illustrated in Table~\ref{tabIII} and plotted in Fig.~\ref{fig15} as a function of beam energy. 

In Fig.~\ref{fig15}, we observe that $c_s^2$ increases monotonically as a function of beam energy for both light and heavy strange hadrons. This may be attributed to the fact that the average temperature of the fireball increases with beam energy which is reflected as the effective temperature dependence of extracted $c_s^2$ \cite{Biswas:2019wtp}. This effect can also be observed in the relative hierarchy in the values of $c_s^2$ for light and heavy strange hadrons. Since heavy strange hadrons freeze-out at higher temperature, the average temperature experienced by them is larger compared to the light hadrons for same beam energy. This is in accordance with the fitted value of $c_s^2$ which is consistently larger for heavy strange hadrons as shown in Fig.~\ref{fig15}. Moreover, this result is also in line with the expectation from Fig.~\ref{158AGeV} which support a clear existence of a mass dependent hierarchy in thermal freeze-out of hadrons. This is the reason we compare the $c_s^2$ values for light and heavy hadrons in Fig.~\ref{fig15}, even though the quality of simultaneous fit for light hadrons is not very good.

Another interesting quantity which we study is the difference of $\eta_{\rm max}$ which directly affects rapidity spectra. In Fig.~\ref{Delta_LH_HS}, we show its difference for light hadron and heavy strange hadrons, $\Delta\eta_{\rm max}=|\eta_{\rm max}^{\rm LH}-\eta_{\rm max}^{\rm HS}|$ as a function of beam energy. We see that this plot shows non-monotonic behavior as a function of beam energy with a minimum at $E_{\rm Lab}=30$A~GeV. This is indeed quite interesting and deserves further attention. Systematic studies of heavy-ion data collected in the domain $E_{\rm Lab}=(20-160)$A~GeV revealed irregular structures in various observables around $E_{\rm Lab}=30$A~GeV beam energy~\cite{Bleicher:2011jk}, which are believed to be connected to the onset of deconfinement transition at low SPS energies. However to attribute the observed minimum in difference of fitted parameters for light hadrons and heavy strange hadrons to the appearance of deconfinement phase transition, one needs a thorough investigation, which is beyond the scope of the present work. Of course one first needs to ensure that all other factors remain the same in the fit of light hadrons and heavy strange hadrons before one can make any robust claim.

Before we close, one may note that within blast-wave framework, the macroscopic thermodynamic parameters are directly extracted by fitting the certain phase-space density distribution of experimentally measured hadrons. Recently the kinetic freeze-out stage has been explored in central Au+Au collisions at energies ranging from $\sqrt{s_{NN}}=2.4$ GeV to $\sqrt{s_{NN}}=200$ GeV, using the microscopic UrQMD model and the corresponding macroscopic parameters are calculated via coarse graining approach~\cite{Inghirami:2019muf}. Results indicate the kinetic freeze-out as a continuous process, leading to a distribution of the freeze-out parameters at different collision energies. The corresponding average kinetic freeze-out temperatures at different beam energies are higher than those obtained by us in our previous work for bulk hadrons. 


\section{Summary}

In conclusion, we have made an attempt to study the hierarchy in the kinetic freeze-out conditions of different hadrons in central Pb+Pb collisions at different SPS energies using non boost invariant blast wave model. Transverse momentum spectra and rapidity spectra of these hadrons, as available, are fitted simultaneously using an iterative scheme to obtain the $\eta_{max}$, $\langle \beta_{T} \rangle$ and $T_{kin}$ values which explain the data reasonably well. We found a clear mass dependent hierarchy in the fitted kinetic freeze-out parameters. This hierarchy of kinetic freeze-out parameters is expected as the medium induced momentum change of heavy hadrons would be smaller compared to lighter hadrons. Therefore, as the temperature of the fireball decreases, one would expect an earlier kinetic decoupling of heavy hadrons. The results indicate that $T_{kin}$ values are in the range $90-110$ MeV with $\langle\beta_{T}\rangle$ of about $0.4c-0.5c$. The temperature values are rather higher than the light particles discussed in~\cite{Rode:2018hlj} which indicate early thermal freeze-out of heavy strange hadrons. The extracted $\eta_{max}$ also explains the corresponding rapidity spectra reasonably well. We found that the extracted freeze-out parameters for charmed hadrons also corroborates this mass dependent hierarchy. The values of $\eta_{max}$, $\langle \beta_{T} \rangle$ and $T_{kin}$ were found to increase monotonously as a function of beam energy.

Moreover, the rapidity spectra of light hadrons as well as heavy strange hadrons are tested with a different model prescription than blast-wave in order to explore the longitudinal properties of the medium. For this, a non-conformal Landau hydrodynamical model description of rapidity distributions from a recent work~\cite{Biswas:2019wtp} is used. This prediction explains the heavy strange hadrons spectra reasonably well however is not satisfactory for light hadrons. We found that the fitted value of sound velocity in the medium also exhibit a similar hierarchy which is obtained from fits to $p_T$-spectra. We advocate that our findings are essential to provide predictions for upcoming experiments at FAIR and NICA accelerator facilities.

Looking forward, it will be interesting to repeat this exercise with charmed hadrons for lower energy collisions, when the data become available. This would be possible with the future measurements at SPS. As mentioned earlier, the NA60$+$ experiment~\cite{Agnello:2018evr} at SPS aims at the measurement of charmonia in Pb+Pb collisons in the beam energy range $E_{lab} = 20A - 158A$ GeV. In addition, the upgraded version of NA61/SHINE experiment at SPS plans to measure the open charm mesons ($D$ meson) via their hadronic decay channel, in Pb+Pb collisions at beam energies $40A$ and  $150A$ GeV~\cite{staszel}. A large statistics data set at $150A$ GeV has already been collected which is presently being analyzed. The existence of mass hierarchy in kinetic decoupling at low energy collisions, can be tested more robustly, if in addition to charmonia, transverse spectra of $D$ mesons are also made available, since their rest mass is closer to that of multi-strange hadrons. We leave this analysis for future.

%
\begin{acknowledgements}
We thank Sushant Singh and Sandeep Chatterjee for critically reading the manuscript. P.P.B. is grateful to Indranil Das and Kaushik Banerjee for stimulating discussions. A.J. is supported in part by the DST-INSPIRE faculty award under Grant No. DST/INSPIRE/04/2017/000038.
\end{acknowledgements}
%

\end{document}